\documentclass[hyper(*)]{JHEP3}

\usepackage{epsfig,multicol,bbm}
\usepackage{enumerate}
\usepackage{bibentry}
\usepackage{amsmath}

\title{Aspects of line operators of class ${\cal S}$ theories}

\author{Dan Xie

\\ School of Natural Sciences, Institute for Advanced Study \\
Princeton, NJ 08540, USA}

\abstract{Geometric picture of line operators of $\mathcal{N}=2$ class ${\cal S}$ theories was found by imposing  closure condition on operator product expansion (OPE) of line operators.
In this paper, we first identify the geometric representation  of ordinary Wilson-'t Hooft line operators of field theory, and study duality action on them.
We further define a Dirac product between line operators and classify the allowed set of  line operators by requiring: a: closure of OPE; b: mutual locality; c: maximality.
Using above classifications, we find many distinct gauge theories associated with a single duality frame, and show explicitly that new possibilities correspond to the choice of 
global form of gauge group and discrete $\theta$ angles. 
 We also study S and T duality actions relating those theories. In particular, we find very interesting duality webs for Maldacena-Nunez 
theory. }

\begin{document}

\section{Introduction}
Wilson and 't Hooft line operators are important extended physical observables for gauge theory: they can be used to probe
 the phases of  field theory;  the study of duality actions on line operators reveals many important property of 
dualities, etc. 

We are interested in studying line operators of four dimensional $\mathcal{N}=2$ class ${\cal S}$ theory which includes Lagrangian theory and non-Lagrangian theory such as  Argyres-Douglas theory\cite{Argyres:1995jj,Argyres:1995xn}.
The study of of this class of theories can provide invaluable insights into the dynamics of quantum field theory. 

$\mathcal{N}=2$ class ${\cal S}$ theory is defined by compactifying 6d $(2,0)$ theory on a Riemann surface with various co-dimensional two defects \cite{Gaiotto:2009we,Gaiotto:2009hg,Xie:2012hs,Chacaltana:2012zy}. 
One of really remarkable thing about this type of construction is that various highly non-trivial properties of  field theory is mapped to the study of
simple geometric objects  on Riemann surface. For instance, the construction of new theories becomes the study of local punctures; duality actions 
of field theory \cite{Argyres:2007cn}  are simply the mapping class group actions of Riemann surface \cite{Gaiotto:2009we}, etc.   

For our interest, line operator of class ${\cal S}$ theory defined using 
$A_1$ theory is related to the closed curves on Riemann surface \cite{Drukker:2009tz,Gaiotto:2010be}, which is a natural set of objects  on which mapping class group action acts \cite{penner2006probing}. 
Things become even more interesting if we consider higher rank theory. Using closure of operator product expansion (OPE) defined between the line operators, 
it is found that the web structure is needed for geometric representation of general line operators of higher rank theory \cite{Xie:2013lca} \footnote{See \cite{Drukker:2010jp,Gomis:2010kv,Ito:2011ea,Gang:2012yr,Cirafici:2013bha,Cordova:2013bza,Tachikawa:2013hya} for other aspects of line operators of class ${\cal S}$ theory.}. 

\begin{center}
\begin{figure}[htbp]
\small
\centering
\includegraphics[width=12cm]{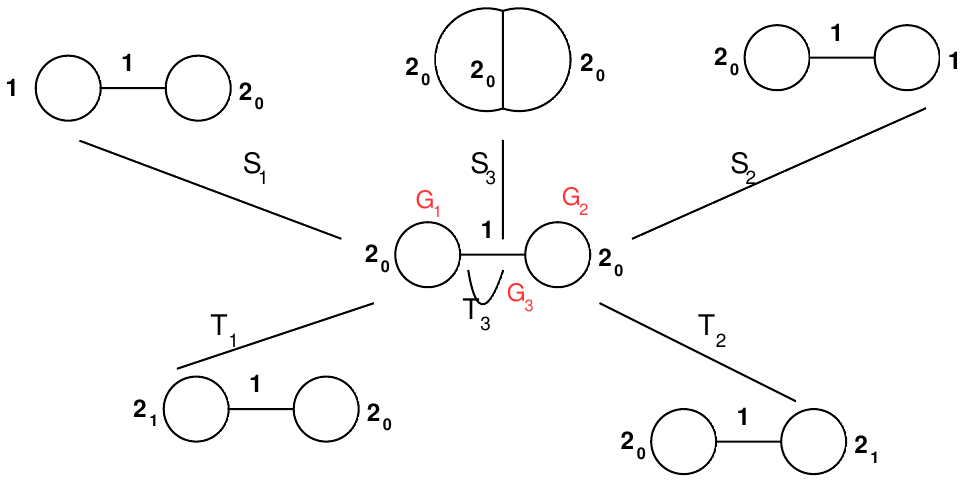}
\caption{Duality action relating different gauge theory defined using genus two Riemann surface. Here $\textbf{1}$ means that gauge group is $SU(2)$, $2_0$ means the gauge group is $SU(2)/Z_2$, and 
$2_1$ means the gauge group is also $SU(2)/Z_2$, but there is a discrete $\theta$ angle associated with it. $S_i$ is the S duality transformation on gauge group and $T_i$ is the $T$ transformation. }
\label{intro}
\end{figure}
\end{center}

In this paper, we further study  line operators of class ${\cal S}$ theory using the geometric picture developed in \cite{Xie:2013lca}, and our main results are summarized as follows:
\begin{itemize}
\item  Geometric objects corresponding to general  Wilson-'t Hooft line operators found in \cite{Kapustin:2005py} are identified. Such correspondence is achieved by 
taking a weakly coupled duality frame and  using Dehn-Thurston coordinates for the closed curves. The crucial thing is that 
web structure is needed to describe general Wilson-'t Hooft line operators.
We give the formula for $T$ and $S$ transformation on line operators represented by closed curves.
 
\item  We define a Dirac product and mutual locality condition between line operators using cluster coordinates. We also give a 
simple geometric interpretations of Dirac product using intersection form of the corresponding homology class, 
and mutual locality condition becomes a constraint on homology class whose study is much more simpler than line operator itself which 
is classified by homotopy theory.

\item We classify the allowed set of line operators by imposing following three conditions: a: closure of OPE; b: mutual locality; c: maximality. 
We identify the corresponding global form of the gauge group and discrete $\theta$ angles. The effect of those discrete $\theta$ angles on line operators are 
studied in a beautiful paper \cite{Aharony:2013hda} generalizing earlier analysis \cite{Gaiotto:2010be}, our results are a further generalization  to 
theory with more gauge groups with fundamental matter (see also \cite{Tachikawa:2013hya} for a similar study of these $\theta$ angles using 6d theory.). 

 \item We study the duality action on homology class, and then relate theories with different  gauge groups and $\theta$ angles. 
 See figure. \ref{intro} for an example.
 \end{itemize}

This paper is summarized as follows: in section 2, we review the geometric picture of line operators found in \cite{Xie:2013lca}, and the important objects are closed 
curves associated with homotopy class and the webs. In section 3, we identify 
the Wilson-'t Hooft line operators in field theory with the geometric objects reviewed in section 2. Section 4 discusses the 
definition of  Dirac product and the importance of intersection form of homology class.  Section 5 focuses on the classification of allowed set of line operators using homology theory, and we also study the duality actions relating 
different theories. Finally, a short conclusion is given.

\section{Closure of OPE: web structure }

\subsection{Geometric picture}
Let's first review the geometric construction of half-BPS line operators of  $\mathcal{N}=2$ class ${\cal S}$  theory found in \cite{Xie:2013lca} (see \cite{Drukker:2009tz,Gaiotto:2010be} for the discussion of $A_1$ theory). 
Class ${\cal S}$  theory is  engineered by compactifying six dimensional  $(2,0)$ theory on a Riemann surface $\Sigma$
with regular and irregular singularities,  see \cite{Gaiotto:2009we,Gaiotto:2009hg,Xie:2012hs,Chacaltana:2012zy} for detailed properties of these singularities. In this paper, we only consider 4d theory engineered using
6d $(2,0)$ $A_{N-1}$ type theory, and only consider full regular and irregular punctures.

For our later purpose, the  punctured Riemann surface could be 
replaced by a bordered Riemann surface by replacing irregular singularity with a disc with marked points \cite{Xie:2012jd}, and the number of marked points depend on the specific 
type of irregular singularity. Therefore 
the geometric avatar for our study is a bordered Riemann surface $\Sigma_{(b_1,b_2,\ldots, b_i),n}$ \footnote{In the following, we simply write $\Sigma$ for the bordered Riemann surface.},
 where $n$ is the number of punctures (regular singularity), and $b_1$ 
is the number of marked points on the boundary of the first disc, $b_2$ 
is the number of marked points on the boundary of the second disc, etc.

\begin{center}
\begin{figure}[htbp]
\small
\centering
\includegraphics[width=5cm]{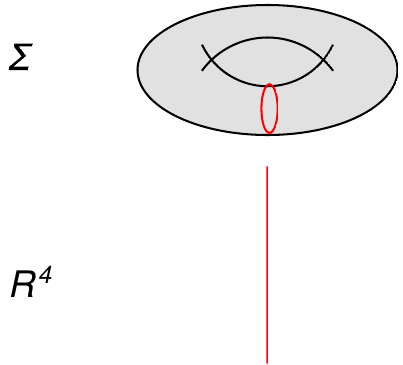}
\caption{A surface operator of 6d $(2,0)$ theory is wrapped on a closed curve on Riemann surface and becomes line operator of 4d theory.}
\label{line}
\end{figure}
\end{center}

Elementary half-BPS line operator of class ${\cal S}$  theory are formed by wrapping half-BPS co-dimension four surface operator of $(2,0)$ theory on closed curves on $\Sigma$. 
These surface operators could be thought of as the boundary of M2 brane ending on M5 brane. After wrapping the surface operator on closed curves of Riemann surface, we get a 
one dimensional object in 4d which is then identified as line operator of four dimensional theory. We simply assume that all line operators in 4d are parallel straight lines, and the 
classification would become the classification of closed curves on $\Sigma$. See figure. \ref{line}.

\begin{center}
\begin{figure}[htbp]
\small
\centering
\includegraphics[width=10cm]{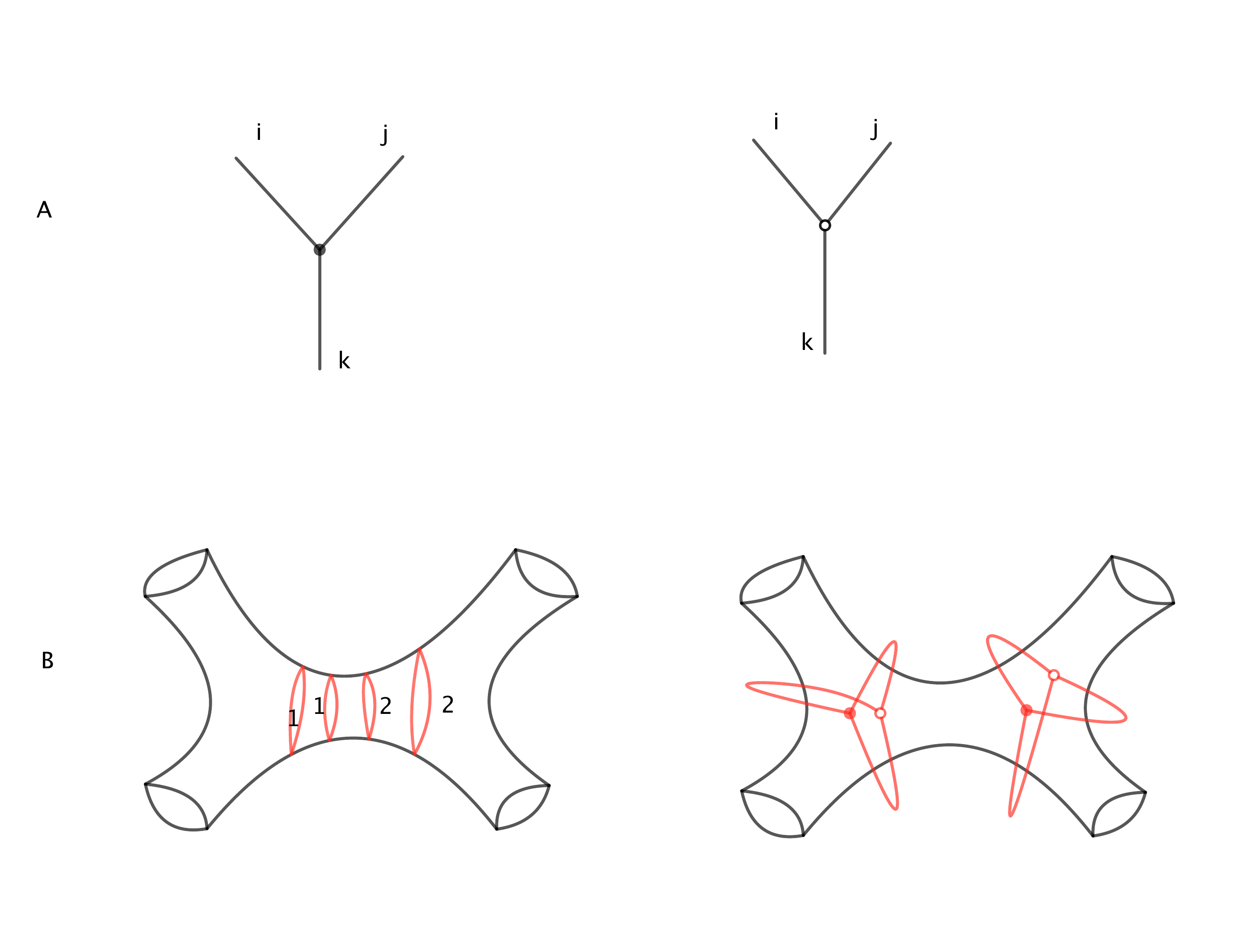}
\caption{Here we give some examples for line operator of $A_2$ theory on a fourth punctured sphere. Left: Wilson line with representation $R=2 w_1+2 w_2$ 
which is represented by colored closed curve. Right: A web operator formed by junctions.}
\label{example}
\end{figure}
\end{center}

 As shown in \cite{Lunin:2007ab, Chen:2007ir, Hoker:2007xz}, half-BPS surface operator is classified by irreducible representation of 
$su(N)$. So if we consider a single closed curve, then the possible 4d line operator one can find from wrapping surface operator is classified by the irreducible representation of $su(N)$ Lie algebra.
The unique highest weight  of an irreducible representation can be expanded as
\begin{equation}
R=\sum n_i \omega_i,
\end{equation}
here $\omega_i$ is the fundamental weight. Motived by this fact, we can first  represent a line operator in fundamental representation $w_i$ by a closed curve with label $i$, 
and  call these labeled curves as colored closed curves. Then we can represent a general line operator in representation $R$ 
by a set of  colored closed curves, and the multiplicity for color $i$ curves is $n_i$, see left of figure. \ref{example}.

A general line operator formed by half-BPS surface operator is then represented by a set of non-intersecting colored closed curves with positive integer weights, and
 those geometric objects are called colored A lamination: 

\textbf{Definition}: A integral \textbf{colored} A-lamination on a bordered Riemann surface is a homotopy class of a collection of 
\textbf{finite} number of self- and mutually nonintersecting \textbf{colored} unoriented curves either closed or connecting two points of the boundary disjoint from marked points with \textbf{integral} weights and subject to the following conditions and equivalence relations:

(1) Weights of all colored curves are positive, unless a curve is special \footnote{Special curve means the curve around the puncture or marked points on the boundary, which can be identified with the flavor line operator.}.

(2) A lamination containing a curve of weight zero is considered to be equivalent to the lamination with this curve removed.

(3) A lamination containing a contractible curve is considered to be equivalent to the lamination with this curve removed.
 
(4) A lamination containing two homotopy equivalent curves  with same color $i$ and weights $u$ and $v$  is equivalent to the lamination 
with one of these curves removed and with the weight $u + v$ on the other.

For $A_1$ theory, the above line operators exhaust all the possibility and agree perfectly with field theory results, see \cite{Drukker:2009tz}. For the higher rank theory, the above 
set of line operator is not complete, and there are new objects we need to consider. 

These new objects are found by studying operator product expansion (OPE) of above line operators: line operator should form a closed algebra in doing OPE.
By calculating OPE explicitly, the web structure for 4d line operators are found. 
Such webs are built by  three junctions labeled by three positive integers satisfying the condition $i+j+k=N$, and vertices could be labeled by black or white. 
One can also form $n$ junction as long as the sum of labels on the external legs is equal to $N$: $i_1+i_2+\ldots+i_n=N$.
In the $A_2$ case, there is only three junction and three labels have to be equal to one, and we can ignore the labels. 
One can form bipartite webs using the above black and white junctions: leg of black and white junction can be connected if they have the same labels, see a web in figure. \ref{example}.
There are equivalence relations between different webs, and line operator is defined by webs modulo equivalence relation. 

In summary,  line operator of 4d theory is represented by following geometric objects 
\begin{itemize}
\item Colored A lamination.
\item Bipartite webs modulo equivalence relation.
\end{itemize}

\subsection{Tropical label for line operators and OPE}
We can label the line operator from colored A lamination by irreducible representation of $su(N)$ lie algebra, but we do not have a good label for line operator from webs. 
There is a tropical label which is applicable to all the line operators considered above. Here we simply review the basic concepts and the interested reader can find more details in \cite{Gaiotto:2010be,Xie:2013lca,Cordova:2013bza}.

These tropical labels are defined by using a triangulation of $\Sigma$ and define a quiver \cite{Xie:2012dw,Xie:2012jd}, and the total 
number of quiver nodes are 
\begin{equation}
r=2n_r+n_f,
\end{equation}
where $n_r$ is the number of Coulomb branch dimensions of the field theory and $n_f$ is the number of mass parameters.  Tropical $a$ coordinates are 
simply a set of discrete numbers defined on the quiver nodes. 
For a line operator in fundamental (defining) representation $w_1$, 
one could find its tropical $a$ coordinates by calculating the trace of its monodromy around the closed curve using the rule found in \cite{fock-2003}.

To do the calculation explicitly, we need to choose a orientation of Riemann surface, which  will induce an orientation on the closed curves, and we take the line operator wrapping once along this oriented closed curve as the line operator in $w_1$. This is the rule used in \cite{Xie:2013lca}. The other colored closed curves  have the same orientation as $w_1$.  We have the following equivalence relation about the orientation and label: changing the orientation of the closed curve and 
change the label from $i\rightarrow N-i$ would denote a same line operator, and using this rule, we can replace all the colored closed curves by oriented curves with label less than $[N/2]$. 
In the $A_2$ case, we can use the above equivalence to replace the colored closed curves with oriented curves without any label. 
The above rule also induces an orientation on junctions: the orientation of black junction is coming out, while the orientation of white junction is coming in. Again, we can change the orientation and label simultaneously 
to put all the labels of the leg to be less than $[N/2]$.

The result from monodromy calculation is always a positive Laurent polynomial in cluster $X$ coordinates, which is denoted as $I(L)$ and called canonical map. This canonical map 
simply means that we represent a line operator by a Laurent polynomial!
 The tropical $a$ coordinate is simply the exponent of the leading order term, which is 
a set of positive integers with fraction ${1\over N}$. For the details on the calculation and examples, see \cite{Xie:2013lca}.

Using the canonical map, one can calculate OPE \footnote{See \cite{Kapustin:2006pk,Kapustin:2007wm,Saulina:2011qr,Moraru:2012nu} for related study of OPE of $\mathcal{N}=4$ theory. Our 
OPE is closed related to theirs but not quite the same.} of two line operators by simply multiplying two Laurent polynomials, and OPE has the following familiar form: 
\begin{equation}
I(L_1)I(L_2)=\sum_L C_{L_1 L_2}^L I(L),
\end{equation}
here $ C_{L_1 L_2}^L$ is a positive integer. The leading order term has tropical $a$ coordinates $a(L_1)+a(L_2)$, and the coefficient $C_{L_1 L_2}^{L_1+L_2}=1$.
In fact, the web structure reviewed in last subsection is found by calculating the OPE explicitly. 

\begin{center}
\begin{figure}[htbp]
\small
\centering
\includegraphics[width=6cm]{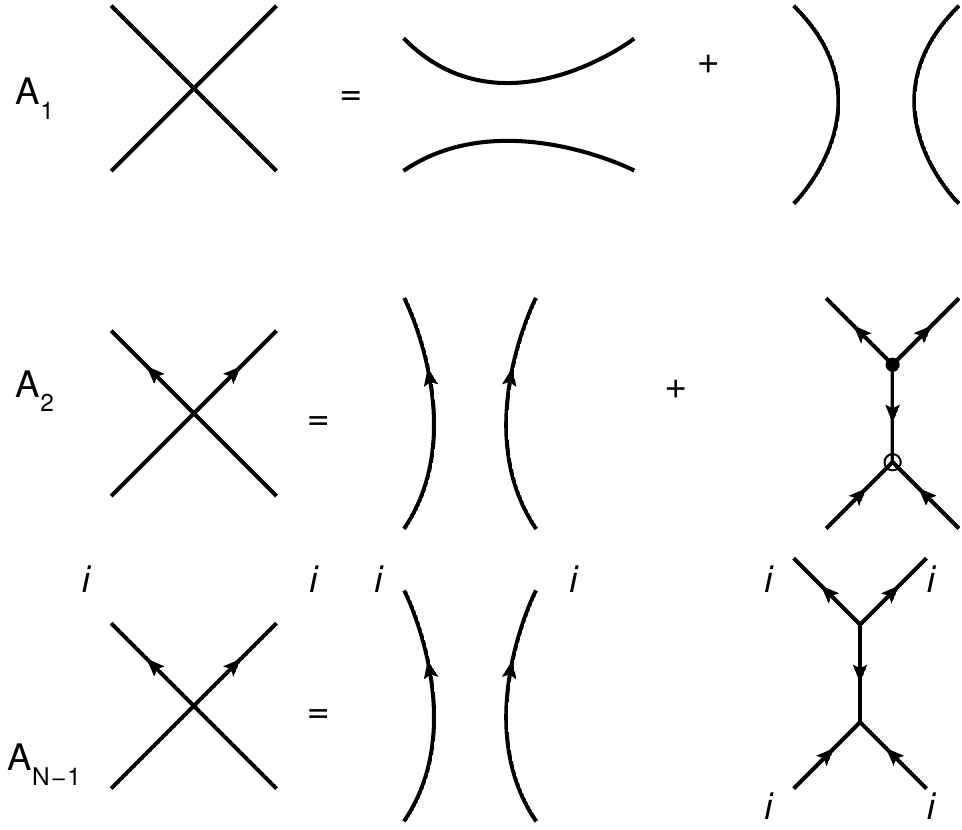}
\caption{ Skein relations which can be used to find OPE in a simple way. Here we only show some simple examples without showing the full set of skein relations.}
\label{skein}
\end{figure}
\end{center}

\begin{center}
\begin{figure}[htbp]
\small
\centering
\includegraphics[width=8cm]{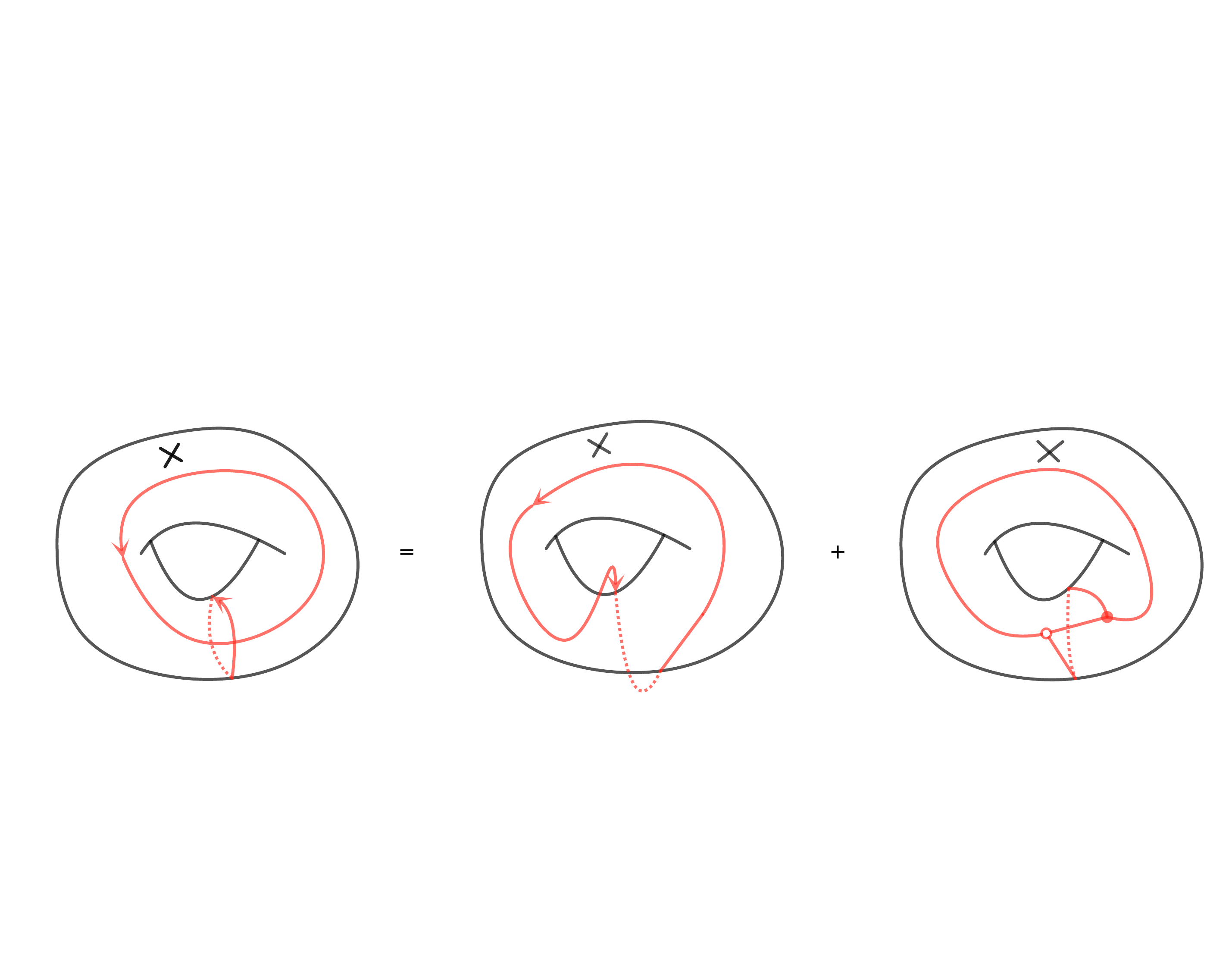}
\caption{The use of skein relation in finding OPE of line operators of $A_2$ theory on a once punctured torus.}
\label{ope}
\end{figure}
\end{center}

Instead of using canonical map, OPE can also be 
calculated using remarkably simple skein relations \cite{Xie:2013lca}, see figure. \ref{skein} for several simple examples. The procedure of finding OPE geometrically is following: first draw two line operators
which will intersect each other, and then use  skein relation to resolve the intersection. See figure. \ref{ope} for an example.

\newpage
\section{Gauge theory interpretation and duality action on line operators}
In this section, we will identify the geometric representation of the familiar Wilson-'t Hooft line operators in field theory and study the duality actions on them.
We only consider theory defined Riemann surface with regular punctures, and use the un-oriented picture of line operators.

\subsection{Classification of Wilson-'t Hooft line operators from field theory}
It is shown by Kapustin \cite{Kapustin:2005py} that a general Wilson-'t Hooft line operator of $\mathcal{N}=4$ SYM is classified by a pair of weights modulo the action of Weyl group:
\begin{equation}
(B, \mu)/W,
\end{equation}
here $B$ is a magnetic weight of dual group $G^L$ and $\mu$ is an electric weight of $G$. For pure electric line operator, the classification is equivalent 
to the classification of irreducible representation  $G$, and similarly pure magnetic line operator is also equivalent to an 
irreducible representation of $G^L$. When $B$ and $\mu$ both nonzero, we can first use Weyl group to fix $B$ to be highest weight of 
an irreducible representation. Let's denote the highest weight of fundamental representations of $su(N)$ as $\omega_i$,  then the highest 
weight of an irreducible representation of $su(N)$ can be written as 
\begin{equation}
B=\sum m_i \omega_i,
\end{equation}
with $m_i\geq 0$. After fixing the magnetic weight, the electric weight can take arbitrary values 
\begin{equation}
\mu= \sum n_i \omega_i,
\end{equation}
and $n_i$ can take both positive and negative values.  Here we ignore the constraints on $m_i$ and $n_i$ due to 
the global form of the gauge group. 

In our case, there are also fundamental matter and sometimes the dual gauge group is also $G$. Therefore we claim the most general 
Wilson-'t Hooft line operators for a duality frame with $n$ gauge groups are classified by 
\begin{equation}
(B_1, B_2, \ldots, B_n, \mu_1, \mu_2,\ldots, \mu_n)/W, 
\end{equation}
and $B_i$ and $\mu_i$ can be expanded using the fundamental weights, and again using Weyl transformation we can take $B_i$ to be a highest weight. The coefficients are constrained by several mutual locality 
conditions: a: the line operator should be mutually local with the matter content. b: the line operator should be mutually local with 
each other. These issues will be discussed later, as the results in this section are not affected by these issues.

\subsection{Geometric representation}
Let's now come back to our geometric representation of line operator, and we will identify the geometric object with the line operator from field theory classification. 
To find this identification, we need to take a pants decomposition which represents a weakly coupled duality frame of field theory.
Let's denote our punctured Riemann surface as $\Sigma_{g,n}$, and fix a pants decomposition: there are $2g-2+n$ pair of pants representing $T_N$ theory, and 
$3g-3+n$ $SU(N)$ \footnote{More precisely, we can only say that the lie algebra of the gauge group is $su(N)$, and we will discuss the global form of the gauge group later.} gauge groups represented by simple closed curves $C_i$. It is now easy to find the Wilson loop of gauge group $G_i$ associated with $C_i$: they are 
represented by a set of closed curves around $C_i$, and it is shown in section 2 that  this type of line operator is indeed classified by the irreducible representation of
$su(N)$. Similarly, by considering the line operators supported on dual cycle of $C_i$, we find that the magnetic line operators are also classified with irreducible 
representation of $su(N)$.

\begin{center}
\begin{figure}[htbp]
\small
\centering
\includegraphics[width=6cm]{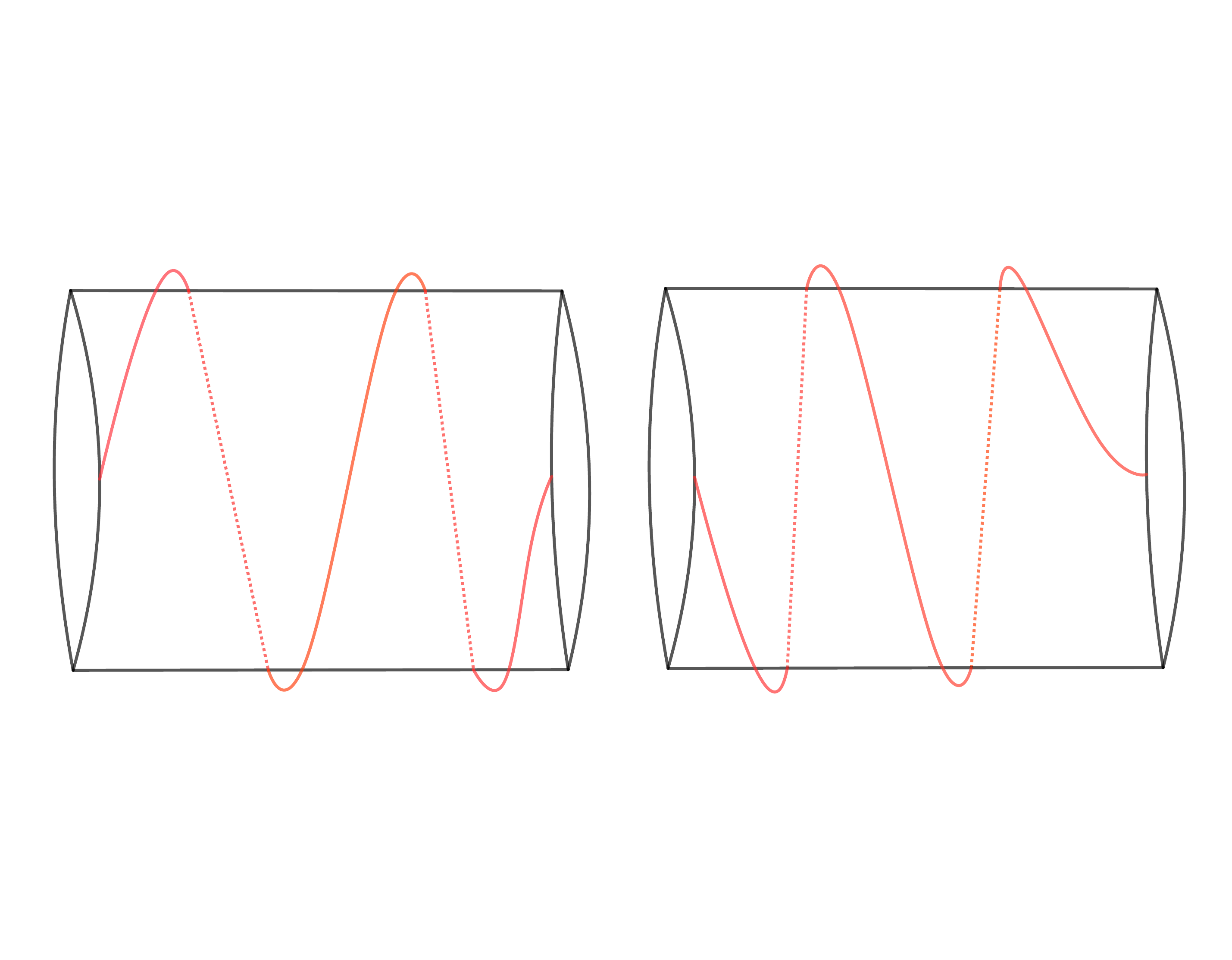}
\caption{Left: the twisting parameter is positive. Right: the twisting parameter is negative. }
\label{twsitco}
\end{figure}
\end{center}

For more general line operator, we need to introduce Dehn-Thurston coordinates for unoriented closed curves (colors will be introduced later).  
Given a pants decomposition, the Dehn-Thurston coordinates for a multiple curve $l$ are a set of integers
\begin{equation}
(m_1, m_2, \ldots, m_{3g-3+n}, t_1, t_2, \ldots, t_n). 
\end{equation}
Here $m_i$ is the geometric intersection number of $l$ and $C_i$, 
and $t_i$ is the twisting number along the annulus region around $C_i$. $m_i$ is positive, and $t_i$ can be positive or negative based on its winding direction (if $m_i=0$, then $t_i\geq 0$), see figure. \ref{twsitco}.
Dehn-Thurston coordinates in a pants satisfy the following condition 
\begin{equation}
m_i+m_j+m_k \in \text{even}
\end{equation} 
if $i,j,k$ are in same pants. 

\begin{center}
\begin{figure}[htbp]
\small
\centering
\includegraphics[width=8cm]{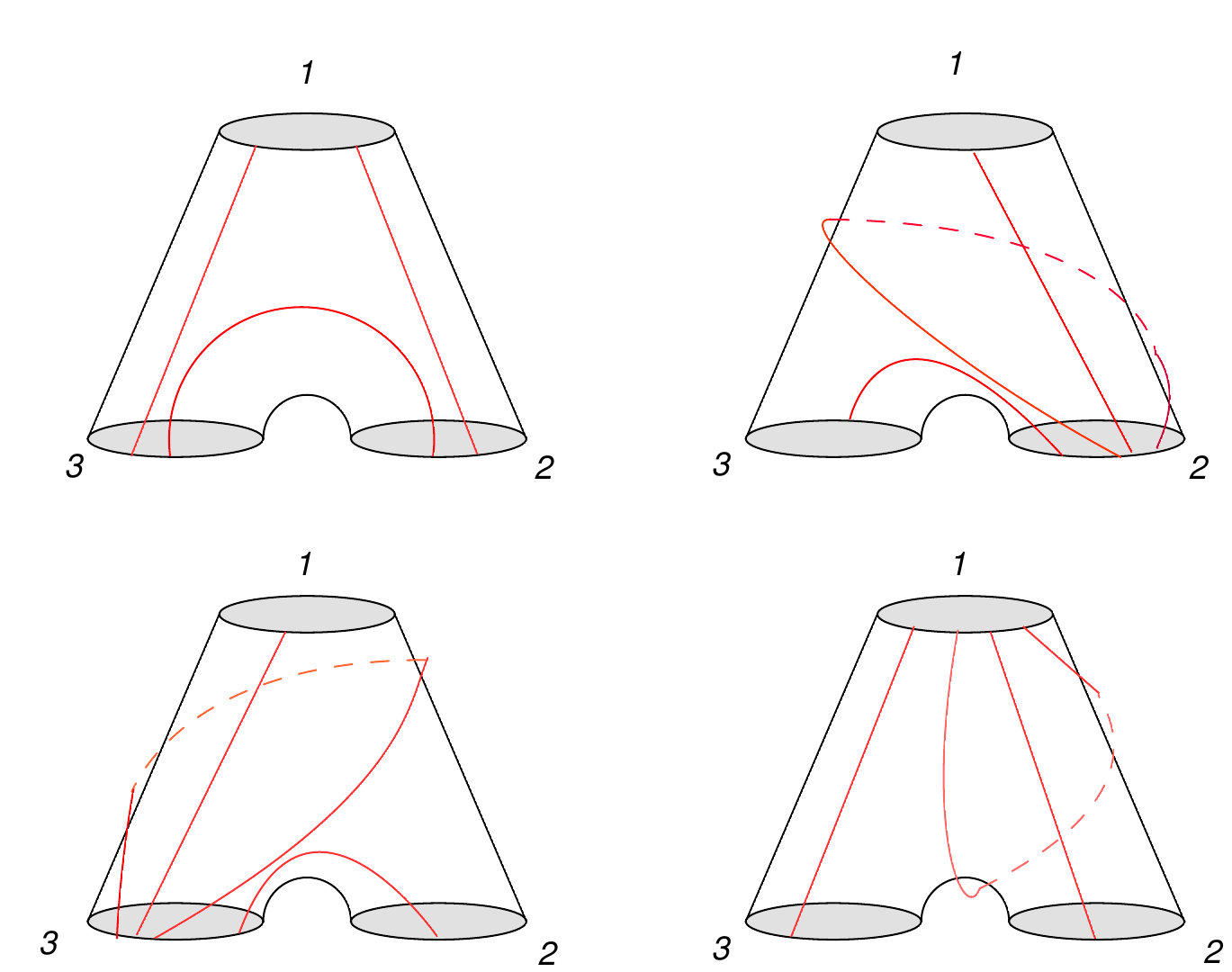}
\caption{Curve models based on different choices of magnetic coordinates on a pants.}
\label{pants}
\end{figure}
\end{center}

\begin{center}
\begin{figure}[htbp]
\small
\centering
\includegraphics[width=8cm]{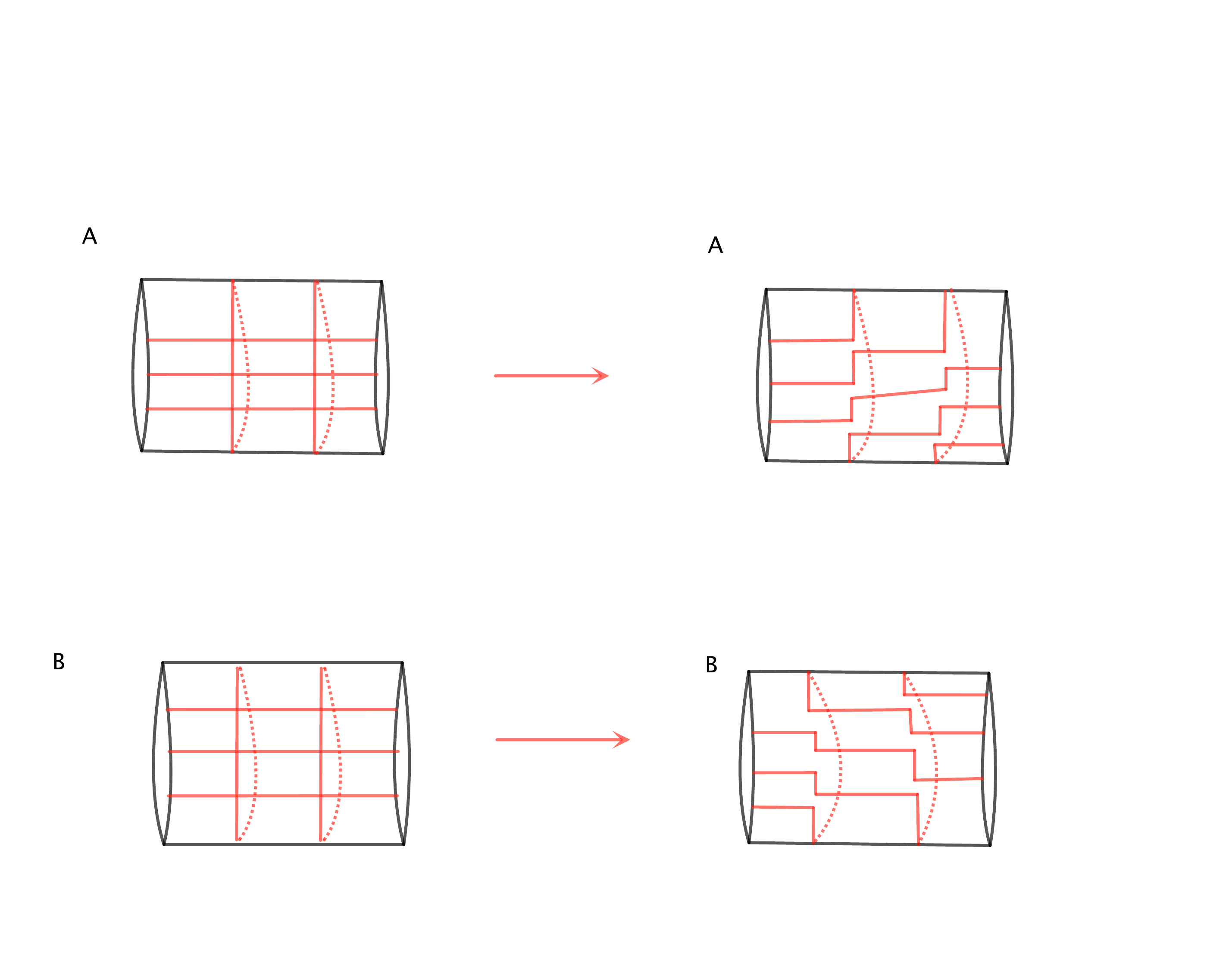}
\caption{One can find the curves from twisting parameter by resolving the intersections in a certain direction which is determined by the sign of twist parameters. }
\label{twist}
\end{figure}
\end{center}

On the other hand, given a set of Dehn-Thurston coordinates satisfying the above condition, we can reconstruct the multiple curves. 
The curves on the pants is constructed using  basic building  block represented by simple curves $l_{12}, l_{13}, l_{23}, l_{11}, l_{22}, l_{33}$,
here $l_{ij}$ denote the curves connecting $i$th and $j$th boundary, and $l_{ii}$ are the open curves connecting the same boundary.  The number of these curves are determined by 
the coordinates $m_i$ and we have 
\begin{align}
& 2l_{ii}=max(m_i-m_j-m_k, 0)\nonumber\\
& 2l_{ij}=max(m_i+m_j-m_k,0)-l_{ii}-l_{jj} \nonumber\\
\end{align}
There are actually four cases based on the relative values of $m_i$, and see figure. \ref{pants} for an illustration.  The construction of  of the curves for twisting parameter $t_i$ is simple, see figure. \ref{twist}:
we use $m_i$ horizontal curves on the tube and use $t_i$ loops around the circle, then we use skein relation of $A_1$ theory to resolve the intersections.  
There are two options in using Skein relations, and the rule is to use one option for all the intersections. This gives the positive and negative windings respectively, see figure. \ref{twist}.

Now let's consider colored A lamination, namely we consider multiple curves with labels, then the Dehn-Thurston coordinates are further partitioned according to 
the colors. Consider a single closed curve $c_i$ in the pants decomposition, and assume the partition from the colors are
\begin{equation}
(m_{1i},m_{2i},\ldots, m_{N-1 i}, t_{1i}, t_{2i},\ldots, t_{N-1 i})
\end{equation}
It is natural to identify this line operator as the Wilson-'t Hooft line operator with label $(B, \mu)$:
\begin{equation}
B_i=m_{1i} w_1+m_{2i} w_2+\ldots+m_{N-1 i}w_{N-1},~~~~\mu_i=t_{1i} w_1+t_{2i} w_2+\ldots+ t_{N-1 i} w_{N-1}
\end{equation}
Notice that all the electric coordinates have the same sign. 

Obviously line operators from colored A lamination do not exhaust the whole set of line operators derived from field theory. The question is: 
What is the geometric representation for  other line operators? The answer is that they are represented by webs. 
To see this, let's consider a theory defined by  $A_2$ theory on  a once punctured  torus. Let's 
take a weakly coupled duality frame by choosing a pants decomposition of once punctured torus, and let's simply take the 
closed circle $a$ as the one defining the weakly coupled gauge group. 
Wilson loops are represented by closed curves around $a$ cycle, and 't Hooft loops are 
represented by closed curves around $b$ cycle. 

\begin{center}
\begin{figure}[htbp]
\small
\centering
\includegraphics[width=10cm]{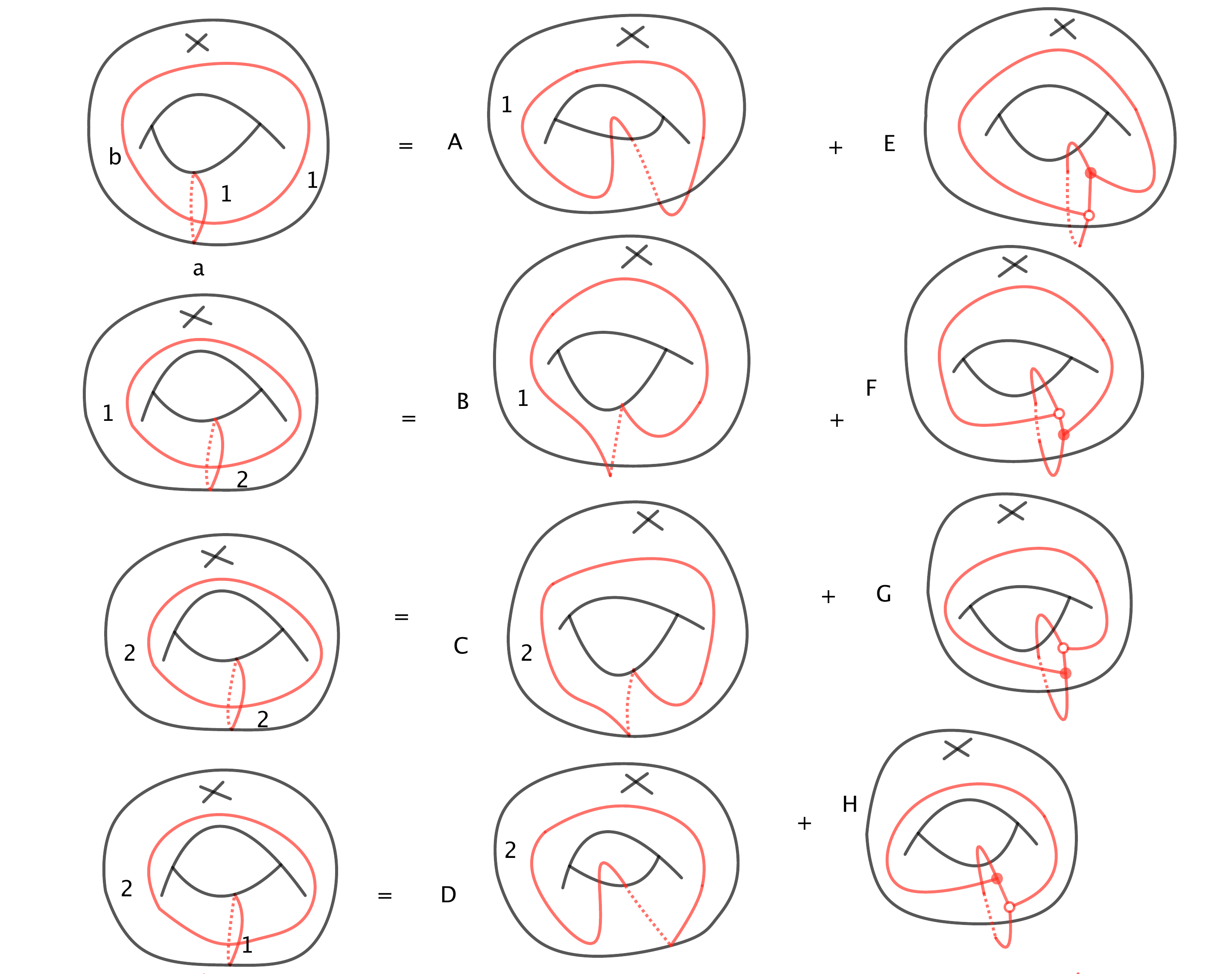}
\caption{Various Wilson-'t Hooft line operators can be found by doing OPE between Wilson and 't Hooft line operators:
A: $\text{WL}_{w_1, w_1}$, B: $\text{WL}_{w_1, -w_1}$, C: $\text{WL}_{w_2, w_2}$, D: $\text{WL}_{w_2, -w_2}$. E: $\text{WL}_{w_1, -w_2}$; F:$\text{WL}_{w_1, w_2}$; G: $\text{WL}_{w_2, w_1}$; H: $\text{WL}_{w_2,- w_1}$ }
\label{general}
\end{figure}
\end{center}

Let's now consider the OPE between the Wilson loops $W_{0, \omega_1}$,  $W_{0, \omega_2}$ 
and 't Hooft line operator $T_{\omega_1,0}$, $T_{\omega_2, 0}$. One can use skein relation to find 
the detailed OPE, see figure. \ref{general}, and we find some line operators represented by the webs. 
It is natural to expect that those webs representing various Wilson-'t Hooft line operators, see figure. \ref{general}, and 
one can find all the missing line operators.  It is not difficult to identify all the Wilson-'t Hooft line operators using 
our OPE construction \cite{xie2014}.  For our consideration of duality action, it is enough to consider line operators
from colored A laminations.
 
\begin{center}
\begin{figure}[htbp]
\small
\centering
\includegraphics[width=4cm]{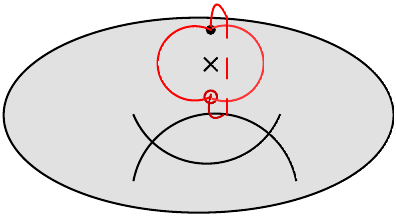}
\caption{New line operator which is not included in Wilson-'t Hooft line operators. }
\label{new}
\end{figure}
\end{center}

One thing we want to stress is that there are new line operators besides the ordinary Wilson-'t Hooft line operators associated 
with gauge group, see figure. \ref{new}. These new line operators can be interpreted as the line operators of $T_N$ theory, and 
they have rather important applications to the construction of Hamiltonian of the underlying integrable system \cite{xie2014}.

\newpage
\subsection{Duality action on line operators}
\begin{center}
\begin{figure}[htbp]
\small
\centering
\includegraphics[width=7cm]{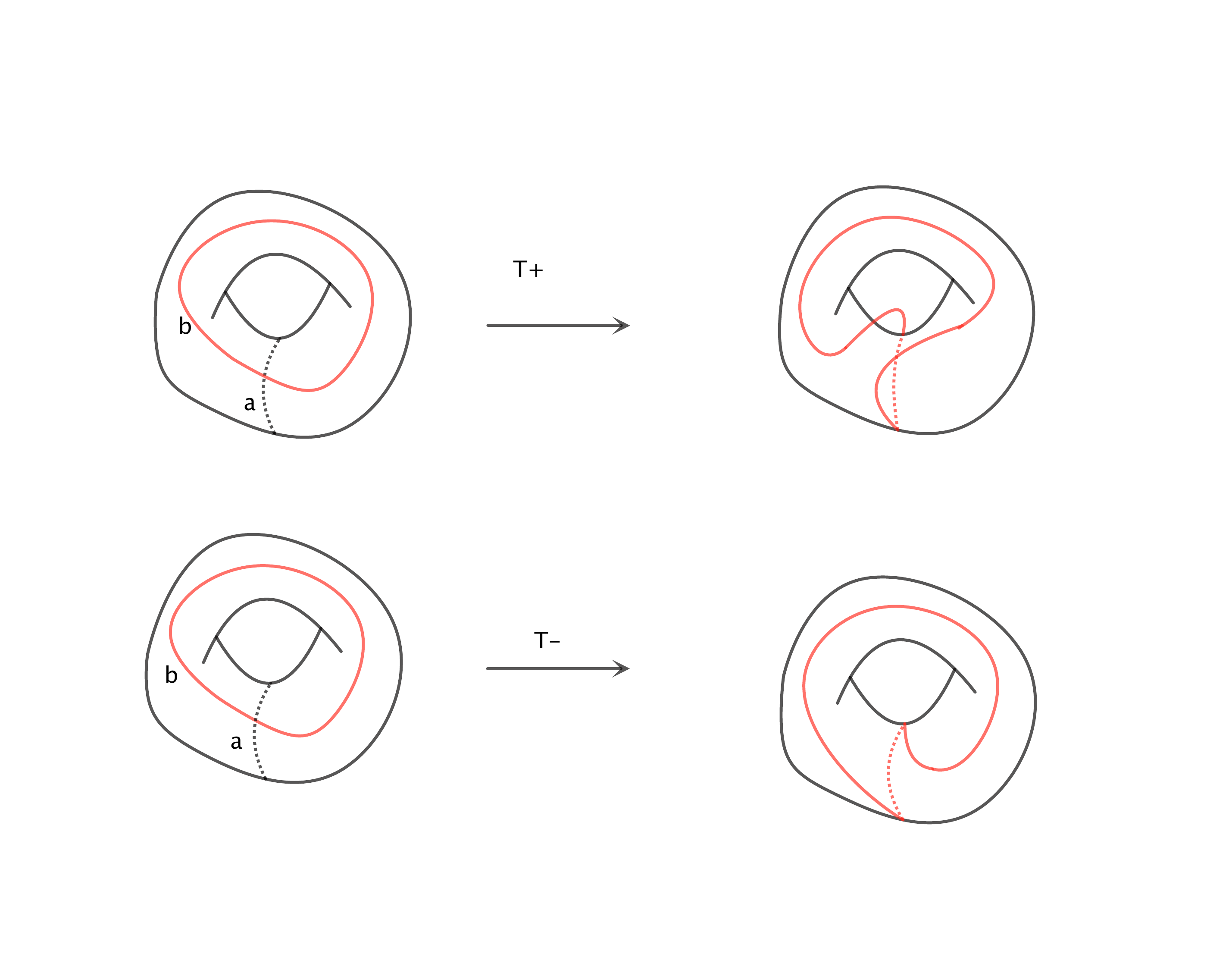}
\caption{Action of $T$ transformation on line operators.}
\label{transform}
\end{figure}
\end{center}

The duality group of four dimensional gauge theory is identified with the mapping class group of the underlying Riemann surface, which 
is generated by the Dehn twist. The Dehn twist around a closed curve acts on line operator coming across that line operator as 
from colored A lamination in certain class at circle $i$ as: 
\begin{equation}
(m_{i}^{'}, t_{i}^{'})=(m_{j}, t_{i}\pm m_{i})
\end{equation}
See figure. \ref{transform}. This action can be thought of as shifting the $\theta$ angle of gauge group by $2\pi$, and the change of electric charge 
due to the $\theta$ angle is essentially the Witten effect \cite{witten1979dyons}.
The generalization of this formula to line operators represented by colored A lamination with label $i$ is simple: for a line operator with label around circle $i$
\begin{align}
&B_i=m_{1i} w_1+m_{2i} w_2+\ldots+m_{N-1 i}w_{N-1},~~~~\mu_i=t_{1i} w_1+t_{2i} w_2+\ldots+ t_{N-1 i} w_{N-1} \nonumber\\
& B_i^{'}=m_{1i} w_1+m_{2i} w_2+\ldots+m_{N-1 i}w_{N-1},~~~ \nonumber\\
&\mu_i^{'}=[t_{1i}+\pm m_{1i}] w_1+[t_{2i}\pm m_{2i}] w_2+\ldots+ [t_{N-1 i} \pm m_{N-1 i}]w_{N-1}
\end{align}

\begin{center}
\begin{figure}[htbp]
\small
\centering
\includegraphics[width=10cm]{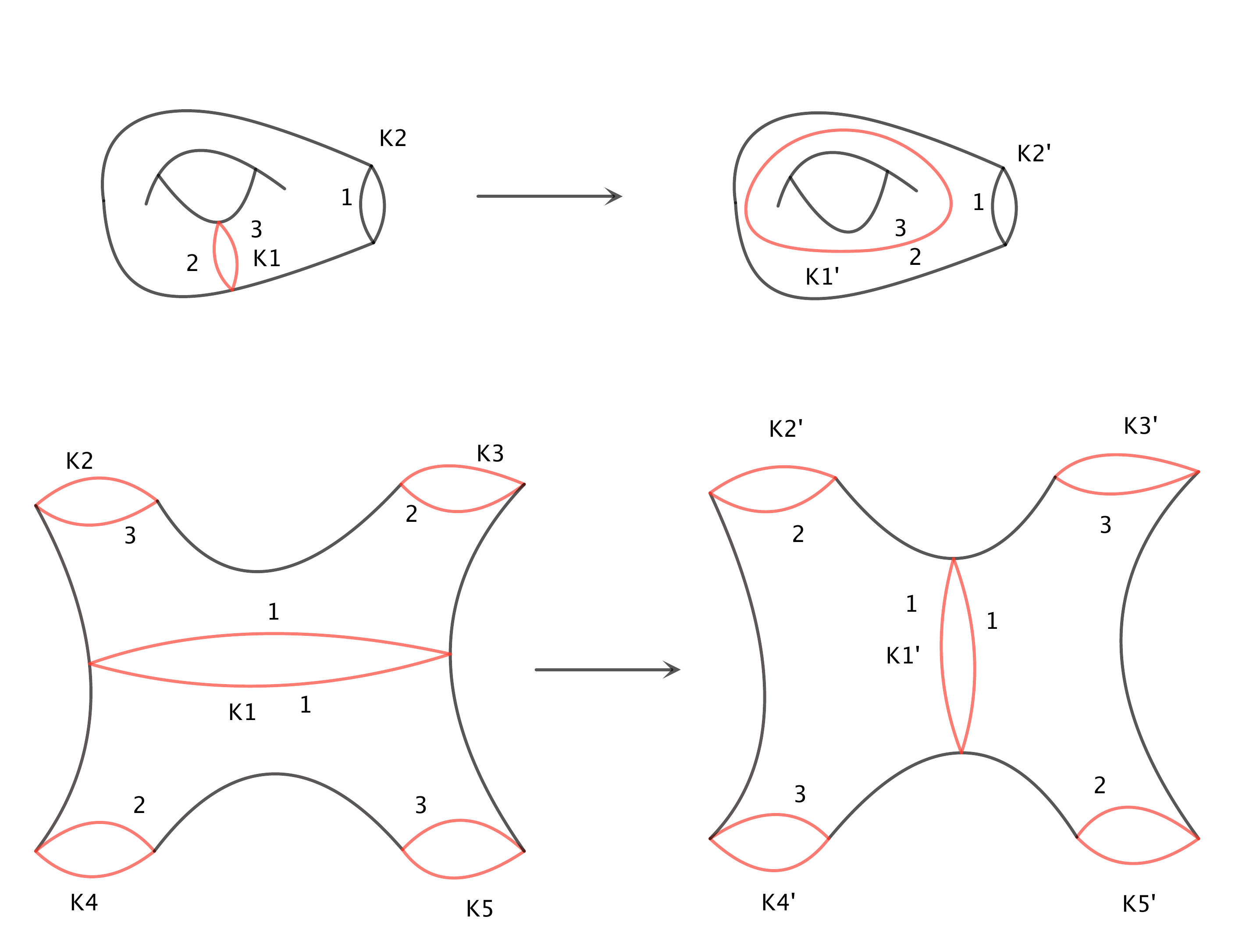}
\caption{Two elementary moves relating pants decomposition. Top: S move relating two pants decomposition of a once punctured torus. Bottom: A move 
relating two pants decomposition of a fourth punctured sphere}
\label{move}
\end{figure}
\end{center}

Now let's fix a pants decomposition, and study the transformation of the Dehn-Thurston coordinates in changing the Pants decomposition. 
It is proven in \cite{hatcher1980presentation} that pants decompositions are related by two fundamental moves: one is the S-move on once punctured torus, and the other one 
is the A move on fourth punctured sphere, see figure. \ref{move}. These two moves are actually the S duality action on the corresponding gauge group. 

The transformation rule on Dehn-Thurston coordinates are found by Penner \cite{penner2006probing}. Since there is only one pants, we have $m_2=m_3$ and $r=\lambda_{12}=\lambda_{13}$. 
 The formula for the S-move on coordinates reads
\begin{align}
&\lambda_{11}^{'}=max(r-|t_1|,0) \nonumber\\
&\lambda^{'}_{12}=\lambda^{'}_{13}=L+\lambda_{11}, \nonumber\\
&\lambda_{23}^{'}=|t_1|-L\nonumber\\
&t_2^{'}=t_2+\lambda_{11}+max(min(L,t_1),0) \nonumber\\
&t_1^{'}=-sgn(t_1)(\lambda_{23}+L)
\end{align}
here $L=r-\lambda_{11}^{'}$. From the above formula, we can easily find the new magnetic coordinates associated with the pants. 

For one application, let's consider line operator with $m_1=0$, and we have $m_2=m_3=m$, and $\lambda_{11}=\lambda_{13}=\lambda_{12}=0,~L=0$,  so we 
can focus locally on a line operator labeled by coordinates $(m,t)$, and the above formula becomes
\begin{align}
&\lambda_{23}^{'}=m^{'}=|t_1|\nonumber\\
&t_1^{'}=-sgn(t_1)m
\end{align}
The generalization of above formula to  line operators of higher rank theory is following 
\begin{align}
&m_{1i}^{'}=|t_{1i}|,~~m_{2i}^{'}=|t_{1i}|,~~\ldots,~~m_{Ni}^{'}=|t_{N-1i}|\nonumber\\
&t_{1i}^{'}=-sgn(t_{1i})m_{1i},~~t_{2i}^{'}=-sgn(t_{2i})m_{2i},~~\ldots,~~t_{N-1i}^{'}=-sgn(t_{N-1i})m_{N-1i},
\end{align}
This formula is derived by only considering the closed curves with the same label, and apply the above formula. The result is the same as found in \cite{Kapustin:2005py}.
The transformation rule for the $A$ move is much more complicated, see \cite{penner2006probing} for the detailed formula. 
The generalization to the higher rank theory is straightforward by considering one type of colored closed curve at one time.

\section{Dirac paring and mutual locality condition}
We used the closure of OPE condition to find the web structure of line operator of class ${\cal S}$ theory. 
In this section, we define Dirac product and mutual locality conditions between line operators, and we use 
the oriented picture for line operators.

The definition of Dirac product is the following: there is a natural Poisson structure on the space of line operators:
\begin{equation}
\{L_1, L_2\}=A I_1+\ldots
\end{equation}
and here $I_1$ is the leading order line operator in the OPE of $L_1$ and $L_2$.  We define  coefficient $a$ as the Dirac product between two line operators as
\begin{equation}
<L_1, L_2>=A,
\end{equation}
and this product is  antisymmetric in exchanging $L_1$ and $L_2$.  Knowing the explicit tropical coordinates of $L_1, L_2$, the constant $A$ has 
an extremely simple form
\begin{equation}
A=\sum \epsilon_{ij} a_{1i} a_{2j}=-\sum x_{1i} a_{2i}
\end{equation}
where $x_{1i}$ is the tropical $x$ coordinates of the line operator $L_1$. 
Two line operators are mutually local if the Dirac product between them are integers:
\begin{equation}
<L_1, L_2>\in  Z.
\end{equation}
The reason this is called mutual locality condition is that this ensures  line operator is not going to pick up a non-trivial phase in going around other line operators.
There  are several useful facts about the Dirac product we defined:
\begin{itemize}
\item In defining tropical $a$ coordinates, we need to choose a coordinate system defined by a quiver on Riemann surface. 
However, it is easy to prove that locality condition is independent of coordinate systems, see appendix for a simple proof.
\item If $L$ is mutually local with a set of line operators $L_i$, then $L_1$ is mutually local with any line operators formed by doing OPE of $L_i$.
The reason is the following: consider the OPE of $L_i$
\begin{equation}
\prod L_j^{n_j}=I_1+C_2 I_2+\ldots,
\end{equation}
here $I_1$ is the leading order term in OPE whose $a$ coordinates are simply the sum of $\sum n_i L_i$, and $I_2$ is the sub-leading order term and $C_2$ is the OPE coefficient.
According to our definition, we have
\begin{equation}
<L, I_1>=\sum n_j<L, L_j>,
\end{equation}
Since $<L, L_j>$ is integer, we see that $<L, I_1>$ is also a integer. Moreover, the difference of tropical $a$ coordinates of $I_i$ and $I_1$
is integer, and $<L, I_j>$ is also a integer for any $j$, so $L$ is mutually local with all the line operators found by doing OPE between $L_i$.
\end{itemize}

Because of the second fact, we only need to study the mutual locality condition on colored closed curves, and consider only web operators from 
doing OPE of mutually local line operators from colored closed curves.

\begin{center}
\begin{figure}[htbp]
\small
\centering
\includegraphics[width=8cm]{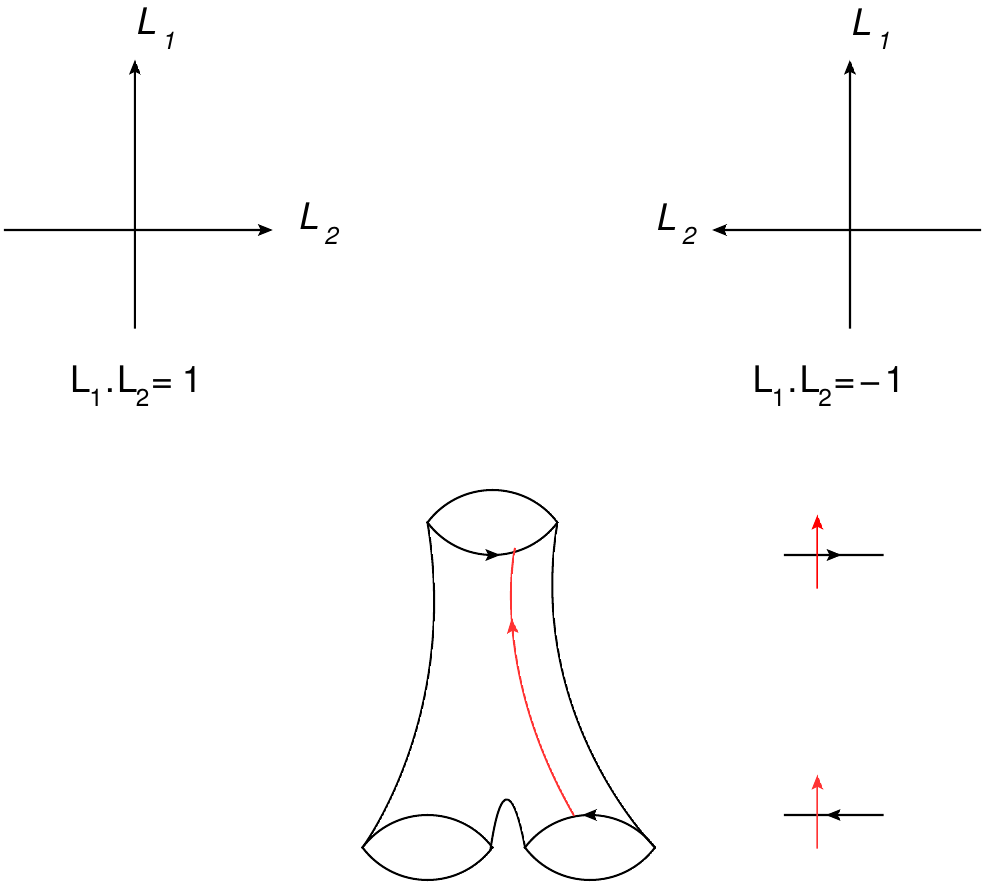}
\caption{Up: Dirac paring for two oriented curves with label $1$. Bottom: a closed curve has opposite contributions to Dirac pairing at two intersections with the boundary of the pants.}
\label{dirac}
\end{figure}
\end{center}

For our application, there is actually a simple graphical rule. Let's take an orientation of 
Riemann surface and calculate the Poisson bracket of two line operators with label $1$, and explicit calculations shows that 
\begin{equation}
L_1\cdot L_2= {ij/N},~~~~\text{or}~~~~L_1\cdot L_2= {-ij/N}
\end{equation}
The sign depends on the orientation, see figure. \ref{dirac}. The Dirac product of two general line operators 
are found by counting the signed intersection number between two set of closed curves.
For later convenience, we multiply Dirac product of two line operators by $N$, and define the mutual locality condition as follows
\begin{equation}
L_1\cdot L_2 \in p N, 
\end{equation}
here $p$ is a integer.

Now here comes the crucial point: 
the above geometric observation of Dirac product tells us that the mutual locality condition depends on signed intersection number\footnote{There is a different kind of intersection number called geometric intersection number which
counts the minimal number of intersections without signs. In $A_1$ case, the mutual locality condition can be defined using either intersection form, but one has to use the signed intersection number for higher rank theory.} (or algebraic intersection number) 
of oriented closed curves, and 
algebraic intersection number is naturally defined on homology class instead of homotopy class. In studying mutual locality condition,
we should work on homology theory instead of homotopy theory in determining the locality between the line operators, which makes the task much easier, as 
the homology theory is abelianization of the homotopy theory.  When we talk about mutual locality condition between line operators, we 
actually mean the corresponding homology class, and each homology class consists many distinct line operators!  

It is easy to find the corresponding homology class of a line operator  $L$ represented by oriented colored closed curve. We first choose a basis of 
homology group and find the corresponding homology class of each oriented closed curve as $[c_j]$, then the corresponding homology class of $L$ is simply
\begin{equation}
[L]=\sum_j i[c_j],
\end{equation}
where $i$ is the color of $c_j$. 

Now let's consider mutual locality condition between line operators and matter represented by three punctured sphere. We define the Dirac pairing as 
the Dirac product between the the oriented line operators and the oriented boundary circles of a pant. Now each curve in the pants has 
a positive and negative contribution to the Dirac product, so their contributions cancel, 
and the locality condition is always obeyed for the line operator from closed curves,
see figure. \ref{dirac}. In $A_1$ case, it is verified in \cite{Drukker:2009tz} that all possible line operators
mutually local with the matter are represented by closed curves. 
It is interesting to verify that all possible Wilson-'t Hooft line operators mutually local with the matter can be represented by 
the line operator considered in last section.

\section{Allowed set of line operators and discrete $\theta$ angle}
Typically, a gauge theory is defined by first specifying the gauge group $G$, and the
allowed matter coupled to gauge group is constrained to transform in representation of  $G$.  After specifying the gauge theory,  
we would like to ask what is the possible choice of line operators. 
The allowed set of line operators are determined as follows (for zero $\theta$ angles):
\begin{itemize}
\item Given the gauge group $G$, the Wilson (electric) line operator is classified by irreducible representation of $G$. 
\item The allowed set of 't Hooft (magnetic) line operators is found by imposing the locality condition with  matter and Wilson line operators. 
\item The allowed set of dyonic line operators is determined by imposing locality condition with electric, magnetic line operators and matter. 
\end{itemize}
We need to include all possible line operators consistent with above conditions. 
There are new possibilities by turning on discrete $\theta$ angles, and the set of line operators are found by doing $T$ transformation on the above 
set of line operators.  So the choice of line operators is not uniquely fixed by  gauge group $G$ and matter content, we need to specify 
discrete $\theta$ angles.

We would like to generalize the above classification of line operators to class ${\cal S}$ theories defined on a genus $g$ Riemann surface with $n$ punctures. In our case, given a duality frame corresponding
to a pants decomposition, the lie algebra of the gauge group is fixed as $g=su(N)^{3g-3+n}$ and the matter content is just $(2g-2+n)$ $T_N$ theory.   
We would like to classify the allowed set of line operators by imposing following four conditions:

\begin{itemize}
\item The line operators are mutually local with the matter content.
\item The line operators form a closed set in doing OPE.
\item The line operators satisfy the mutual locality condition.
\item The sets of line operators form a maximal set, i.e. they are not a subset of any other allowed choice.
\end{itemize}
The first condition implies the identification of line operator (generators) with the colored closed curves on Riemann surface. The second condition
has been used in \cite{Xie:2013lca} to find new line operators represented by webs. We would like to impose  third and fourth conditions on
line operators using the mutual locality conditions defined in last section.

Since web operators are found by doing OPE of operators from colored closed curves, and they will be mutually local to each other if the corresponding 
colored line operators are mutually local, we only need to impose condition on colored closed curve.  
Moreover, the mutual locality condition is a condition on homology class and not the homotopy class, so the task of 
classification is much easier. In the following we always talk about the homology class, and it should be clear that there are many distinct line operators 
in a single homology class. 

\subsection{Gauge group from choice of Wilson line operators}
Usually we first know the gauge group $G$ and matter content, then we try to find the allowed line operators. Here for the class
${\cal S}$ theory, the situation is reversed. We only know the lie algebra, and we want to find the gauge group from 
allowed set of line operators.

Let's choose a pants decomposition of oriented Riemann surface and focus on a gauge group represented by a closed circle $c_i$ whose homology class is denoted $[c_i]$. 
The global form of gauge groups can be found by the allowed homology class of line operators supported on $[c_i]$:
\begin{itemize}
\item If  $[c_i]$ is in trivial homology class or more generally its intersection number with other homology class is zero, then the gauge group is $SU(N)$ as line operators in homology class $[c_i]$ 
is mutually local with all the other line operators, in particular, Wilson loop in defining representation of $su(N)$ is allowed and therefore
the gauge group has to be $SU(N)$.
\item If $[c_i]$ is in a non-trivial homology class, and the minimal allowed homology class is $k[c_i]$ and all the other allowed ones are integer multiple of the minimal one, 
notice that $k$ has to be divided by $N$ (otherwise the choice of line operators are not closed under OPE). 
Then the gauge group is $SU(N)/Z_k$. 
\end{itemize}

\begin{center}
\begin{figure}[htbp]
\small
\centering
\includegraphics[width=5cm]{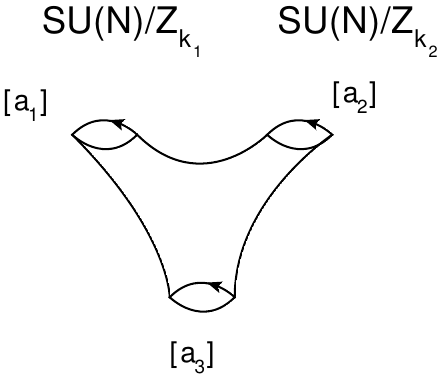}
\caption{The gauge group from choice of Wilson loops on three boundaries of a trinion. }
\label{trinion}
\end{figure}
\end{center}

However, we can not choose the minimal homology classes arbitrarily. To illustrate this point,  
let's consider a pants whose three flavor symmetries are gauged, and we assume that three homology class associated with  boundaries are non-trivial, see figure. \ref{trinion}.
Let's assume that the minimal choice of homology class around
$[a_1]$ and $[a_2]$ are ($k_1[a_1]$, $k_2[a_2]$), then the choice of homology class around $[a_3]$ is fixed, since we have the following relation between homology class:
\begin{equation}
[a_3]=-[a_1]-[a_2],
\end{equation}
and we can find line operators in integer multiple of $[a_3]$ homology class using OPE. It is easy to find what is the minimal homology class of $[a_3]$: 
let's denote the minimal homology class around $[a_3]$ as $p[a_3]$, and $p$ can be found by imposing the condition: 
\begin{equation}
nk_1[a_1]+m k_2 [a_2]=p[a_3]\rightarrow nk_1[a_1]+m k_2 [a_2]=-p[a_1]-p[a_2],
\end{equation}
which implies that $p$ is the minimal number such that
\begin{equation}
n k_1= p,~~~m k_2=p;
\end{equation}
and $p$ is uniquely fixed once $k_1$ and $k_2$ is given. The gauge group associated with $[a_3]$ is $SU(N)/ Z_p$. 

The above consideration only tells us the local (a single gauge group) form of the gauge group, to find out the real form of the gauge group, we need to 
use the matter content which is described by $T_N$ theory.  

$T_N$ theory is defined by compactifying 6d theory on a sphere with three full punctures, and the lie algebra of flavor symmetry is 
\begin{equation}
su(N)_a\times su(N)_b\times su(N)_c,
\end{equation}
and naively the center would be $\Gamma=Z_N^{a}\times Z_N^{b} \times Z_N^{c}$, however,  as pointed out in \cite{Tachikawa:2013hya}, the true center is actually not $\Gamma$ but just a single $Z_N$.  
Assuming the generator of $Z_N^{a},Z_N^{b},Z_N^{c}$ as $\gamma_1, \gamma_2, \gamma_3$, then the above statement of a single $Z_N$ center of $T_N$ theory 
can be interpreted as relations between the three generators:
\begin{equation}
\gamma_1\gamma_2^{-1}=1,~~~~~\gamma_2\gamma_3^{-1}=1.
\label{relation}
\end{equation}
$T_N$ theory has a set of operators $Q_{abc}$ which transforms as trifundamental representation of the flavor group, and therefore is not invariant under the center $Z_N$. 
This puts constraints on possible global form of gauge group:  $Q_{abc}$ should be invariant under ungauged center of the flavor group. 
If only one flavor group is gauged, then the gauge group has to be $SU(N)$, and if only two flavor groups are gauged, then the gauge group has to be $SU(N)/(Z_k)_1 \times SU(N)/(Z_k)_2$, 
and the generators of  $(Z_k)_1$ and $(Z_k)_2$ are $\gamma_1^{N/k}$ and $\gamma_2^{-N/k}$. Due to relation [\ref{relation}], $T_N$ theory is invariant under the ungauged center. 
In general, if three flavor groups of $T_N$ theory are gauged, and we have following choices \footnote{This part is motivated by the  very helpful discussions with K.Yonekura.}:
\begin{itemize}
\item  the gauge group is  $SU(N)\times SU(N)\times SU(N)$: $Z_N$ is gauged, and there is no residual discrete flavor symmetry.
\item  the gauge group is $ SU(N)/Z_k \times SU(N)/Z_k \times SU(N)$, and the generators of two $Z_k$ are $\gamma_1^{N/k}$ and $\gamma_2^{-N/k}$.
\item the gauge group is $ SU(N)/Z_{k} \times SU(N)/Z_{k} \times SU(N)/Z_{k}$. The generators for three ungauged center are $(\gamma_1^{N/k_1},\gamma_2^{-N/k_1}, \gamma_3^{N/k_1})$.  
\item the gauge group is $ SU(N)/Z_{k_1} \times SU(N)/Z_{k_2} \times SU(N)/Z_{k_3}$. To ensure $T_N$ theory invariant under the ungauged center, one can not choose $k_i$ arbitrarily.  Let's assume 
that we fix $k_1$ and $k_2$ with $k_1\neq k_2$, and $k_1$ and $k_2$ should be divided by N. The generators for first two ungauged center are $(\gamma_1^{N/k_1}, \gamma_2^{-N/k_2})$.  Using relation [\ref{relation}], this two center generate 
a group $Z_{p}$ whose generators is $\gamma_2^{N/ p}$. $p$ is determined by finding the minimal $p$ such that 
\begin{equation}
n k_1=p,~~~m k_2 =p; 
\end{equation}
Here $n$ and $m$ are integers. To make sure $T_N$ theory  invariant under the ungauged center, we must have 
\begin{equation}
k_3=p,
\end{equation}
and the third gauge group has to be $SU(N)/Z_p$. This constraint is precisely the one we found using the homology class of pants!
\end{itemize}

\subsection{Action of $T$ and $S$ transformation on homology class}
We can use S and T transformation to relate different gauge theories. S duality is simply interpreted as changing the pants decompositions which could be decomposed into 
two fundamental moves; $T$ transformation corresponds to 
doing Dehn twist around closed circle whose homology class is denoted as $[a]$,  and the action of Dehn twist on homology class $[b]$  is \cite{farb2011primer}:
\begin{equation}
T_a([b])=[b]+\hat{i}(a,b)[a],
\label{T}
\end{equation}
and here $\hat{i}(a,b)$ is the algebraic intersection number between two homology classes, which only depends on homology class $[a]$ and $[b]$.
Physically, $T$ transformation corresponds to change the discrete $\theta$ angle by $2\pi$.

The $S$ duality action on homology class is more complicated. Consider once punctured torus and the $S$ move, 
 $S$ duality action simply exchanges  $[a]$ and $[b]$ cycle. Now to preserve the canonical intersection number,  the action is actually
\begin{equation}
[a^{'}]=[b],~~~[b^{'}]=-[a]
\end{equation}
so the $S$ duality action on a general homology class $[l]=n[a]+m[b]+\ldots$ is 
\begin{equation}
S[l]=m[a^{'}]-n[b^{'}]+\ldots.
\end{equation}
One can also find $S$ duality action on homology class associated with the $A$ move on fourth punctured sphere, for our purpose, the action 
simply exchanges the cycle associated with the original gauge group and the cycle associated with the dual cycle.

\subsection{Examples}
After connecting all the basic tools, 
 we will give various examples in this section on the choice of allowed line operator, and then identify the global form of the gauge groups and discrete $\theta$ angles. 
 Duality webs among these theories are also studied. 

\subsubsection{$T_N$ theory: no choice}
Let's discuss first the allowed set of line operators of $T_N$ theory. 
The basic non-trivial line operators of $T_N$ theory are formed by webs built from two three junctions, see figure. \ref{TN}. Any two web operators
are mutually local, as the two intersections have opposite orientation, and they are mutually local, see figure. \ref{TN}.
Therefore, there is no choice of line operators for $T_N$ theory, which agrees with the conclusion in \cite{Tachikawa:2013hya}. 
\begin{center}
\begin{figure}[htbp]
\small
\centering
\includegraphics[width=8cm]{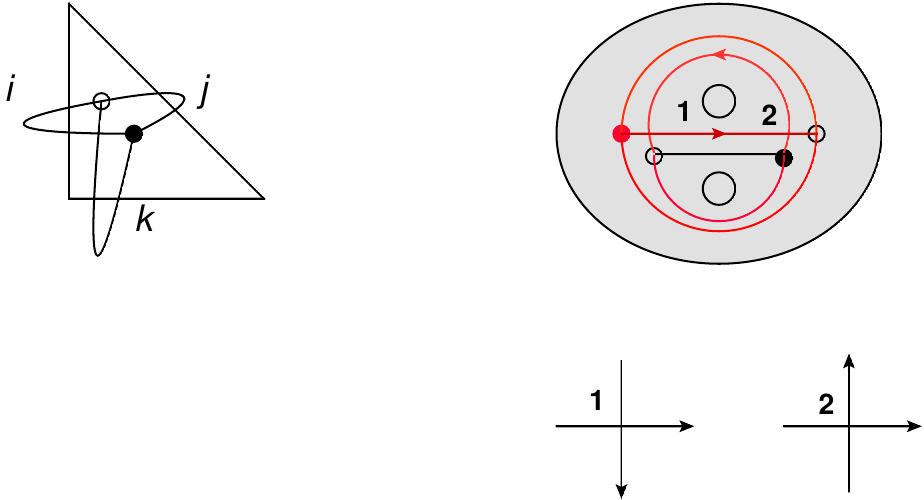}
\caption{Left: the generators for non-trivial line operators of $T_N$ theory. Right: the intersections between two web operators; Two intersections have opposite 
intersection number, so two line operators are mutually local.}
\label{TN}
\end{figure}
\end{center}

\subsubsection{Sphere with punctures: no choice}
\begin{center}
\begin{figure}[htbp]
\small
\centering
\includegraphics[width=4cm]{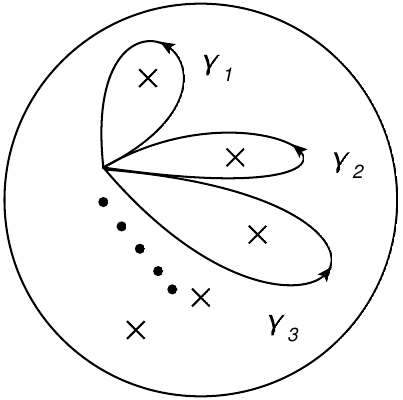}
\caption{The generators of homology class of a punctured sphere.}
\label{sphere}
\end{figure}
\end{center}

The generators for the homology group of punctured sphere are shown in figure. \ref{sphere}, and they satisfy the condition:
\begin{equation}
\gamma_1\gamma_2\ldots \gamma_n=1.
\end{equation}

The algebraic intersection number of these generators are zero, therefore the mutual locality condition is always satisfied for any line operators. 
Due to the maximality condition on the choice of 
line operators, we can not choose a subset, therefore there is a unique choice of line operators associated with the sphere:
all possible line operators considered in \cite{Xie:2013lca} should be included.
When we choose a duality frame, i.e. take a pants decomposition,  then the gauge groups are $SU(N)$ as  basic Wilson line associated with the 
defining representation of $SU(N)$ is allowed. 

\subsubsection{Genus one case}
Let's first consider $\mathcal{N}=2$ theory defined by a  torus with one  regular singularity. The classification of line operators 
is the same as the  $\mathcal{N}=4$  theory discussed  in \cite{Aharony:2013hda}, here we show how that classification can be implemented using our geometric picture (our consideration is 
also applied to $\mathcal{N}=4$ case). 

The non-trivial homology class of torus is generated by two cycles $a$, $b$, and $c$ around the puncture , and the intersection number is given by 
\begin{equation}
[a]\cdot [b]=1.
\end{equation}
\begin{center}
\begin{figure}[htbp]
\small
\centering
\includegraphics[width=6cm]{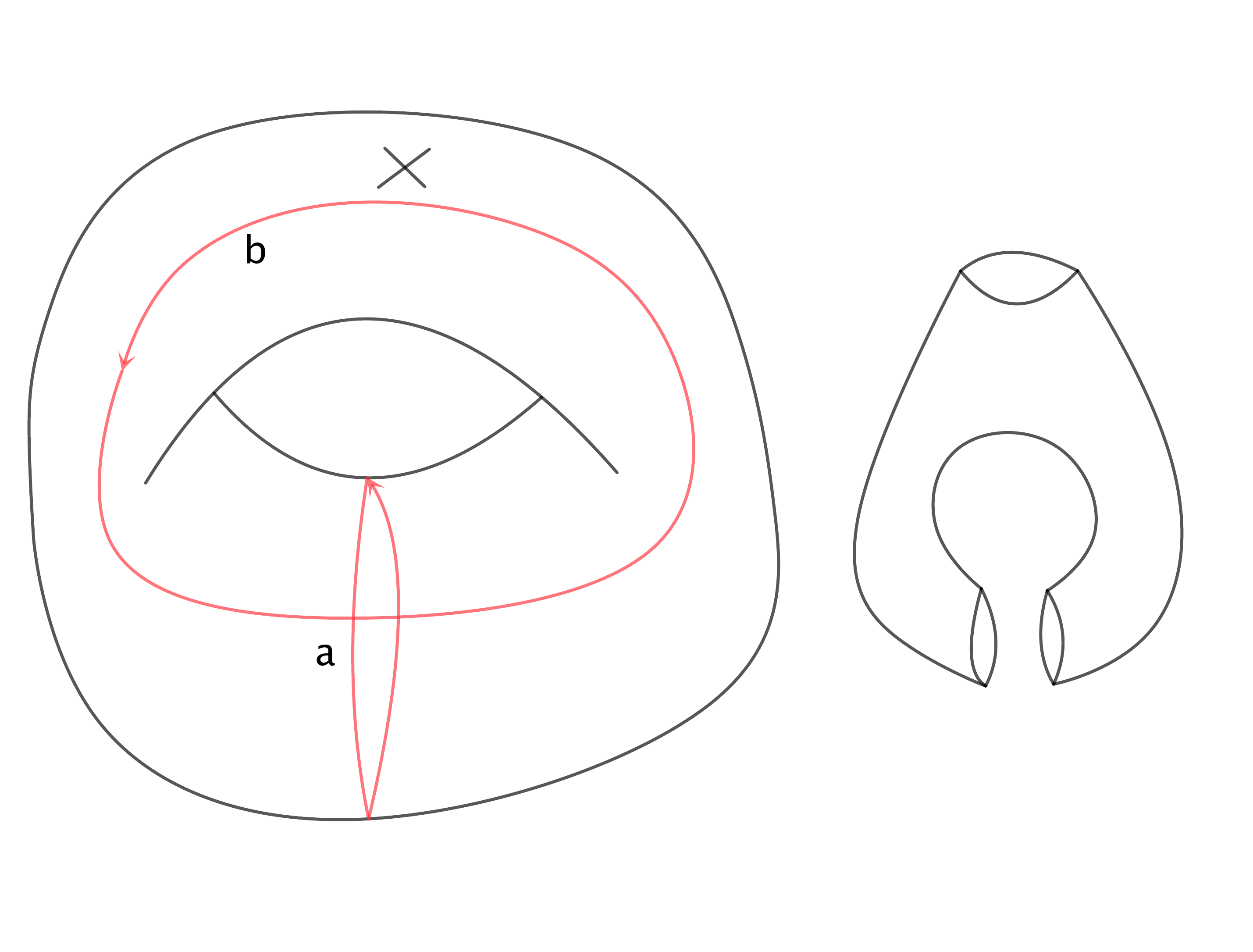}
\caption{Left: the homology class of a once punctured torus. Right: one pants decomposition of once punctured torus. }
\label{torus}
\end{figure}
\end{center}
see figure~\ref{torus}. $c$ has zero intersection numbers with  other two generators, so it can not give any constraints from the consideration of mutual locality condition, and 
we can ignore it from now on. 
Let's take one weakly duality frame which is described by the pants decomposition of the torus, which is defined 
by a closed curve whose homology class can be taken as $[a]$. 

\begin{center}
\begin{figure}[htbp]
\small
\centering
\includegraphics[width=14cm]{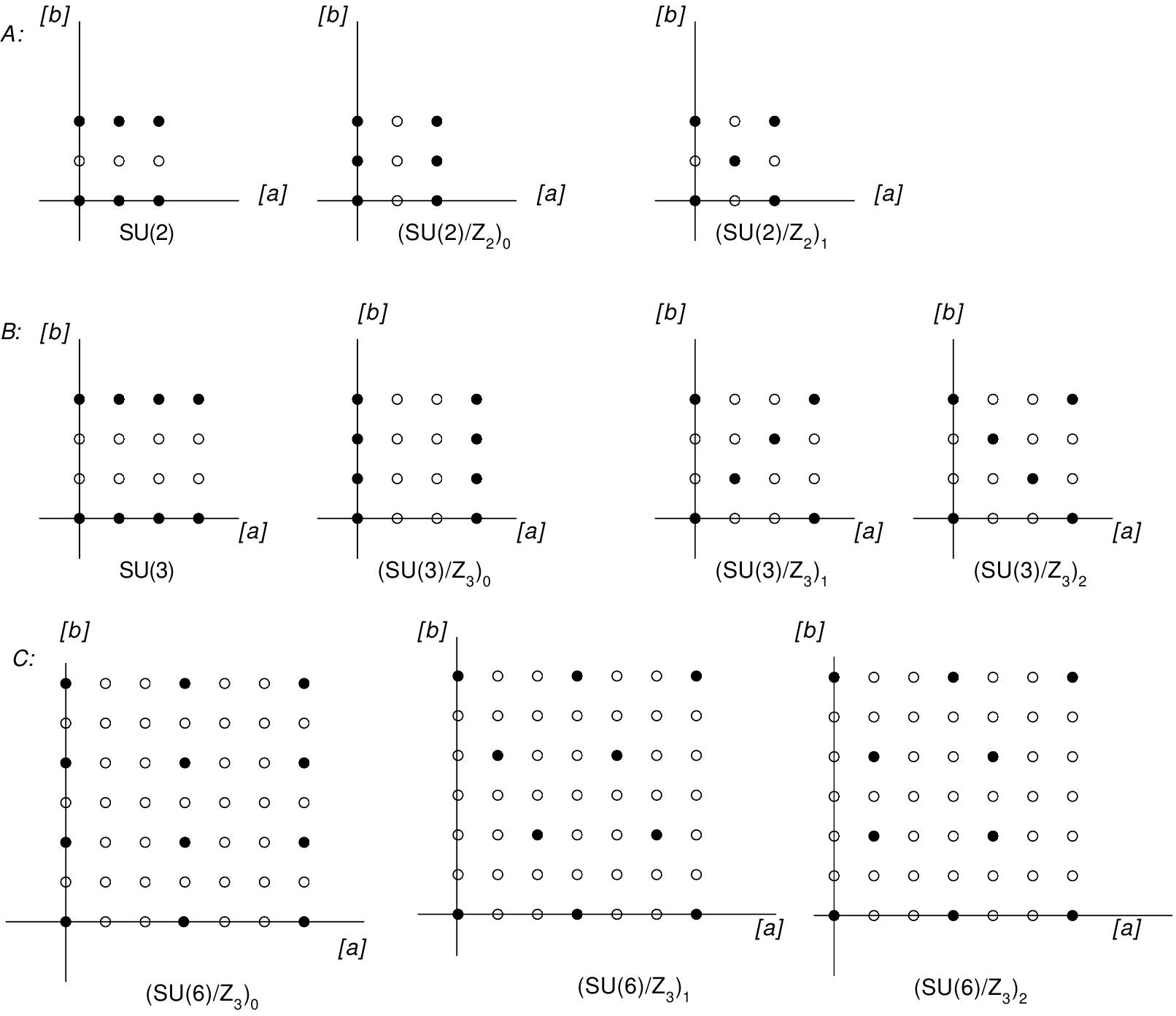}
\caption{Homology class of allowed line operators. }
\label{lattice}
\end{figure}
\end{center}

Let's start with zero $\theta$ angle, and the allowed set of line operators are classified as:
\begin{itemize}
\item The minimal Wilson loop is chosen in homology class $k[a]$ with $k$ divides N, i.e. $k k^{'}=N$. This  implies that the gauge 
group is $SU(N)/Z_k$. The choice of electric line operators are $n[a]$ with
\begin{equation}
n= p_1 k.
\end{equation}
\item The pure magnetic line operators are chosen by imposing the mutual locality condition with  electric line operators. Let's denote the corresponding 
homology class as $m [b]$, and we have
\begin{equation}
m[b]. k[a]=p_2N\rightarrow m=p_2 k^{'},
\end{equation}
\item Let's denote the homology class  of  a general line operator as $n[a]+m[b]$, we require it to be mutually local with the basic electric and magnetic line operators:
\begin{equation}
(n[a]+m[b])\cdot k[a]=p_2N,~~~(n[a]+m[b]) \cdot k^{'}[b]=p_1N
\end{equation}
which implies that $m =p_2 k^{'}$ and $n =p_1 k$. 
\end{itemize}
It is easy to check the mixed line operators determined by above conditions are mutually local to each other. In summary, the homology class of allowed line operators are 
\begin{equation}
p_1 k[a]+p_2 k^{'}[b],
\end{equation}
We can represent the choice of line operators by a lattice in which $x$ ($y$) coordinates are coefficient of $[a]$ ($[b]$), see figure. \ref{lattice}. In fact, it is sufficient to just 
look at the lattice inside the fundamental square bounded by four points $(0,0), (0,N), (N,0), (N,N)$.  The theory defined by above data
is denoted as $(SU(N)/Z_k)_0$.  This type of lattice is essentially the same as the one given in \cite{Aharony:2013hda}, here we interpret them as the allowed homology class of line operators. 

We can find other theories by doing $T$ transformation (turning on discrete $\theta$ angle) around $a$. Let's start with $SU(N)/Z_k$ gauge theory with zero $\theta$ angle, and
do $T$ transformation which corresponds to one Dehn twist around circle $a$.   For a line operator in homology class $[l]=p_1 k [a]+p_2 k^{'} [b]$, the intersection number 
with $[a]$ is $[a]\cdot[l]=p_2 k^{'}$, and using the formula [\ref{T}], the action of Dehn twist  $T_{+}^n$ on homology class $[l]$ is  
\begin{equation}
p_1 k [a]+p_2 k^{'} [b]\rightarrow({p_1 k + n p_2 k^{'}})[a]+p_2 k^{'}[b];
\end{equation}
Using this formula, we see that the choice of Wilson loop is not changed, but the choice of magnetic and mixed  line operator is changed, which will give new theories if 
\begin{equation}
n p_2 k^{'} \neq  k p.
\end{equation}
for any $p_2$. We label these new theories as $(SU(N)/Z_k)_n$ with $n$ means the theory is derived by doing $n$ Dehn twist for the theory with zero $\theta$ angle. 
See figure. \ref{lattice} for some examples, and it is easy to find all possible lattices from doing $T$ transformation on theory with zero $\theta$ angle. There are some simple features about
the $T$ transformation by looking at the above formula:
\begin{itemize}
\item If the gauge group is $SU(N)$ which means that the electric homology class is taking all possible integers, then the $T$ transformation will not change the charge lattice.
\item If the gauge group is $SU(N)/ Z_k$, then the maximal number of non-trivial Dehn-twist is $(k-1)$.  
\item $T$ transformation preserves the mutual locality condition: if two line operators are mutually local before the transformation, they will be mutually local after the transformation.
\end{itemize}
So  we get an allowed set of line operators satisfying our four conditions after  $T$ transformation due to the third feature. 

To identify the dual gauge theory after $S$ transformation,
we need to look at the action of $S$ transformation on homology class:
\begin{equation}
n[a]+m[b]\rightarrow m[a^{'}]-n[b^{'}]
\end{equation}
The action on charge lattice is simple: the new fundamental region is found by rotating 90 degree of the fundamental region in clockwise direction.

Using the above $S$ and $T$ transformation on charge lattice, it is then easy to identify the dual theories.
There are very interesting duality webs as studied in \cite{Aharony:2013hda}. Notice that not all theories can be related by $S$ and $T$ transformations, and 
there are interesting orbit structure.  We only give several simple examples here, the interested reader can work out more complicated example using our geometric representation. 

\begin{center}
\begin{figure}[htbp]
\small
\centering
\includegraphics[width=12cm]{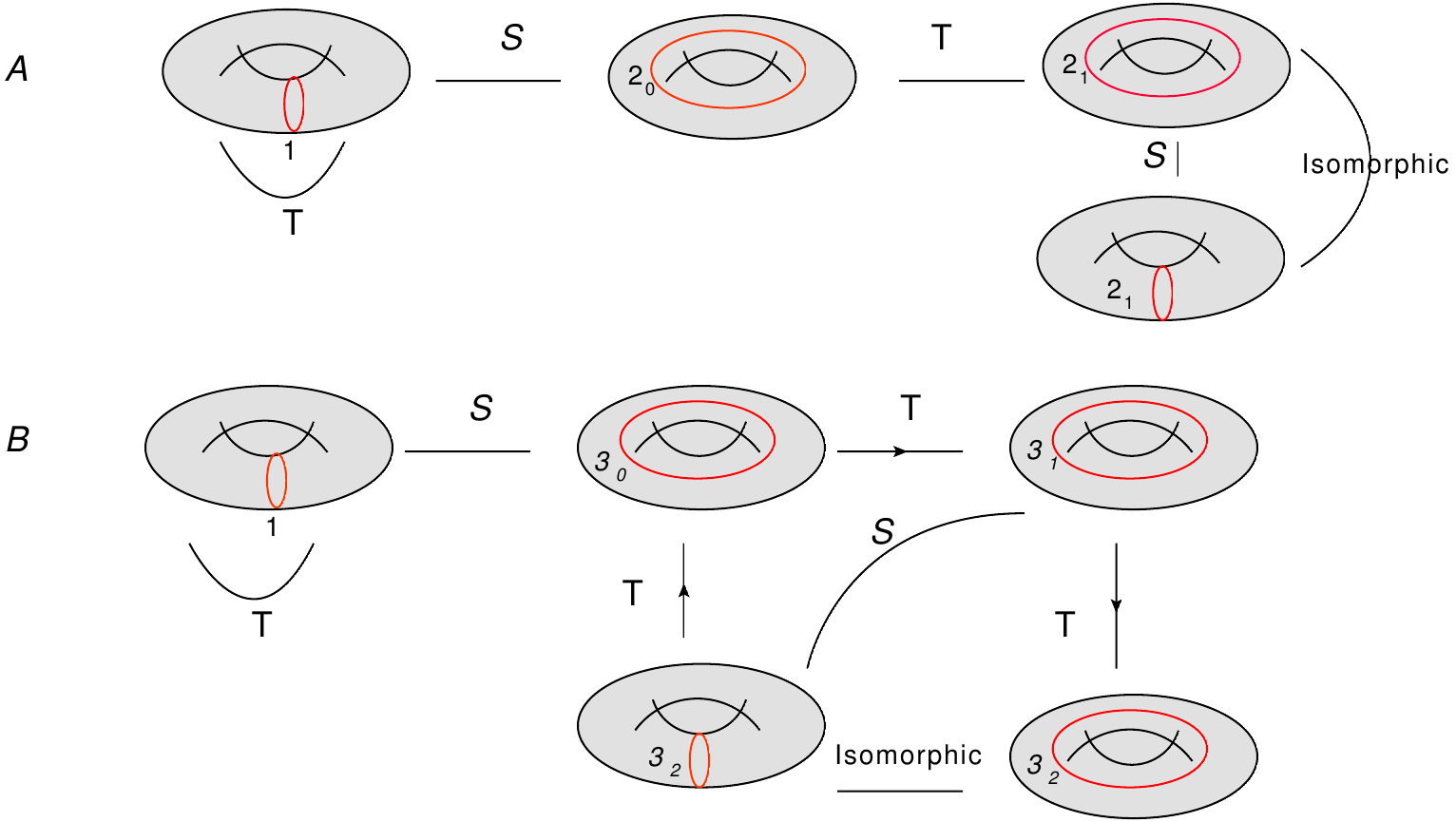}
\caption{Allowed homology class of line operators. }
\label{TN}
\end{figure}
\end{center}

\textbf{Torus with two punctures}:

The homology group of this geometry is generated by $a$ cycle and $b$ cycle of the torus, and  two closed cycles  around 
the punctures. The only nontrivial intersection forms among the generators are 
\begin{equation}
[a]\cdot [b]=1.
\end{equation} 
The classification of the allowed line operators is the same as the above one. Here we want to study the gauge theory 
interpretation and point out one new novelty. 

\begin{center}
\begin{figure}[htbp]
\small
\centering
\includegraphics[width=10cm]{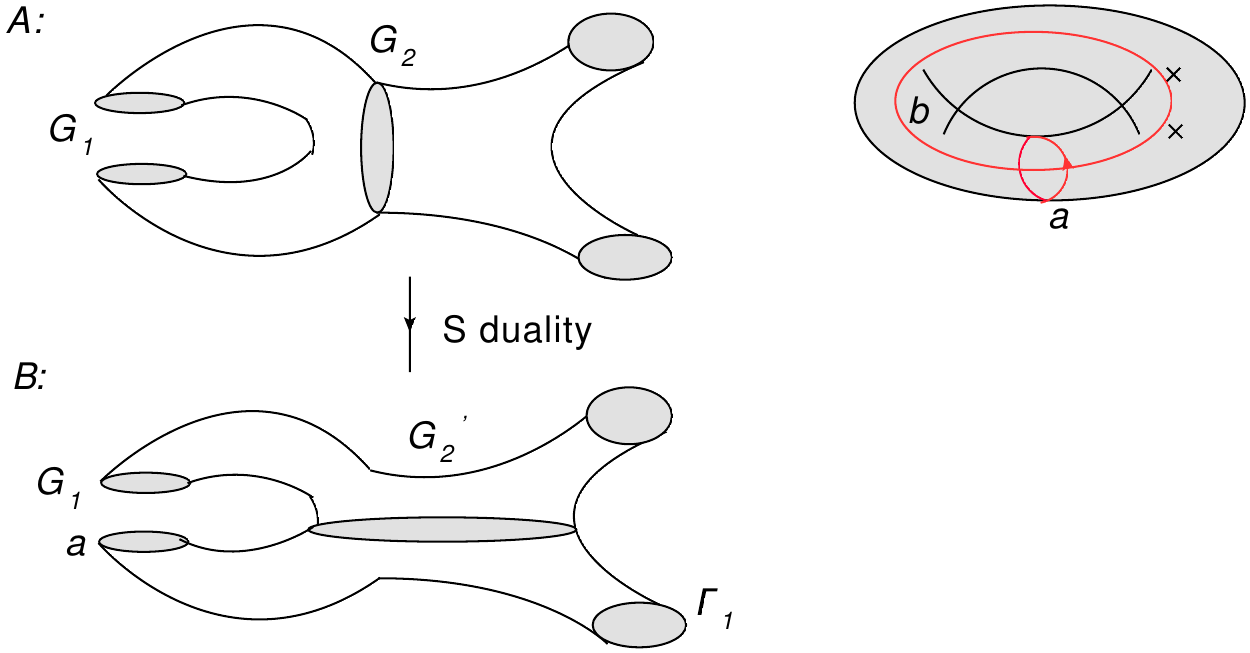}
\caption{S duality action on a theory defined on a torus with two punctures, and the dual gauge group $G_2^{'}$ depends on the choice of gauge group $G_1$.  }
\label{torus2}
\end{figure}
\end{center}

There are two duality frames of this theory: one duality frame has one gauge group coupled to a single $T_N$ theory, and the other gauge group coupled 
to two $T_N$ theories.  In one duality frame shown in figure. \ref{torus2}A, we can choose the minimal homology class as $k[a]$, and 
therefore the gauge group $G_1=SU(N)/Z_k$. 
The closed curve associated with $G_2$ is in trivial homology class, and therefore
the gauge group is $G_2=SU(N)$. The action of $T$ and $S$ duality on $G_1$ group is the same as above case, and we do not repeat the 
analysis here. 

What is interesting is the $S$ and $T$ transformation on gauge group $G_2$.  First of all, $T$ transformation would not change the allowed set 
of line operators. After doing $S$ transformation on group $G_2$, we went to duality frame shown in figure. \ref{torus2}B. Now the 
closed curve associated with $G_2^{'}$ is not in trivial homology class, and it is actually in homology class $[a]+[\Gamma_1]$, since $\Gamma_1$ 
has no intersection with other curves, therefore the winding pattern of Wilson loop on $G_2$ is the same as $G_1$. We conclude that the gauge group is actually
\begin{equation}
SU(N)_1\times SU(N)_2 \over Z_k
\end{equation}
and the $Z_k$ action on two $SU(N)$ groups is generated by $\gamma_1^{N/k}$ and $\gamma_2^{-N/k}$ satisfying the condition 
\begin{equation}
\gamma_1\gamma_2^{-1}=1,
\end{equation}
so the matter is invariant under the ungauged discrete $Z_k$ symmetry. Notice that although locally the gauge group $G_2$ under duality is 
coupled to two $T_N$ theories, its dual gauge group actually depends on the form of $G_1$.

No essentially new things happen if we consider theory defined by a genus one Riemann surface with arbitrary number of punctures.

\subsubsection{Genus two case}
Let's  consider a four dimensional theory defined on a genus two Riemann surface without any puncture, and this class of theory is first studied by
Maldacena and Nunez \cite{Maldacena:2000mw}. The first homology group is generated by $([a_1], [a_2])$ and $([b_1], [b_2])$, and 
the intersection numbers are 
\begin{equation}
[a_1]\cdot [b_1]=1,~~ [a_2]\cdot [b_2]=1.
\end{equation}

Now let's discuss the choice of line operators. First, we need to take a maximal set of non-intersecting basis, and here we take them to be $[a_1]$ and $[a_2]$, then the classification is done as:
\begin{itemize}
\item The choice of electric homology class is the same as genus one case we just studied, and the basis can be denoted as
\begin{equation}
e_1=n_1 k_1[a_1],~~~~~~e_2=n_2 k_2[a_2],
\end{equation}
and here $k_1$ and $k_2$ satisfies the condition $k_1 k_1^{'}=N$ and $k_2 k_2^{'}=N$. 
\item  Let's  denote the homology class of pure magnetic line operator as $m_1[b_1]+m_2[b_2]$, then the mutual locality condition implies that 
\begin{equation}
m_1=p_1 k_1^{'},~~m_2= p_2 k_2^{'}.
\end{equation}

\item For a general homology class $[L]=q_1[a_1]+q_2[a_2]+m_1[b_1]+m_2[b_2]$, mutual locality condition with the electric and magnetic line operators simply implies
\begin{equation}
[L]=n_1 k_1[a_1]+n_2 k_2[a_2]+p_1 k_1^{'}[b_1]+p_2 k_2^{'}[b_2].
\end{equation}
\end{itemize}

\begin{center}
\begin{figure}[htbp]
\small
\centering
\includegraphics[width=10cm]{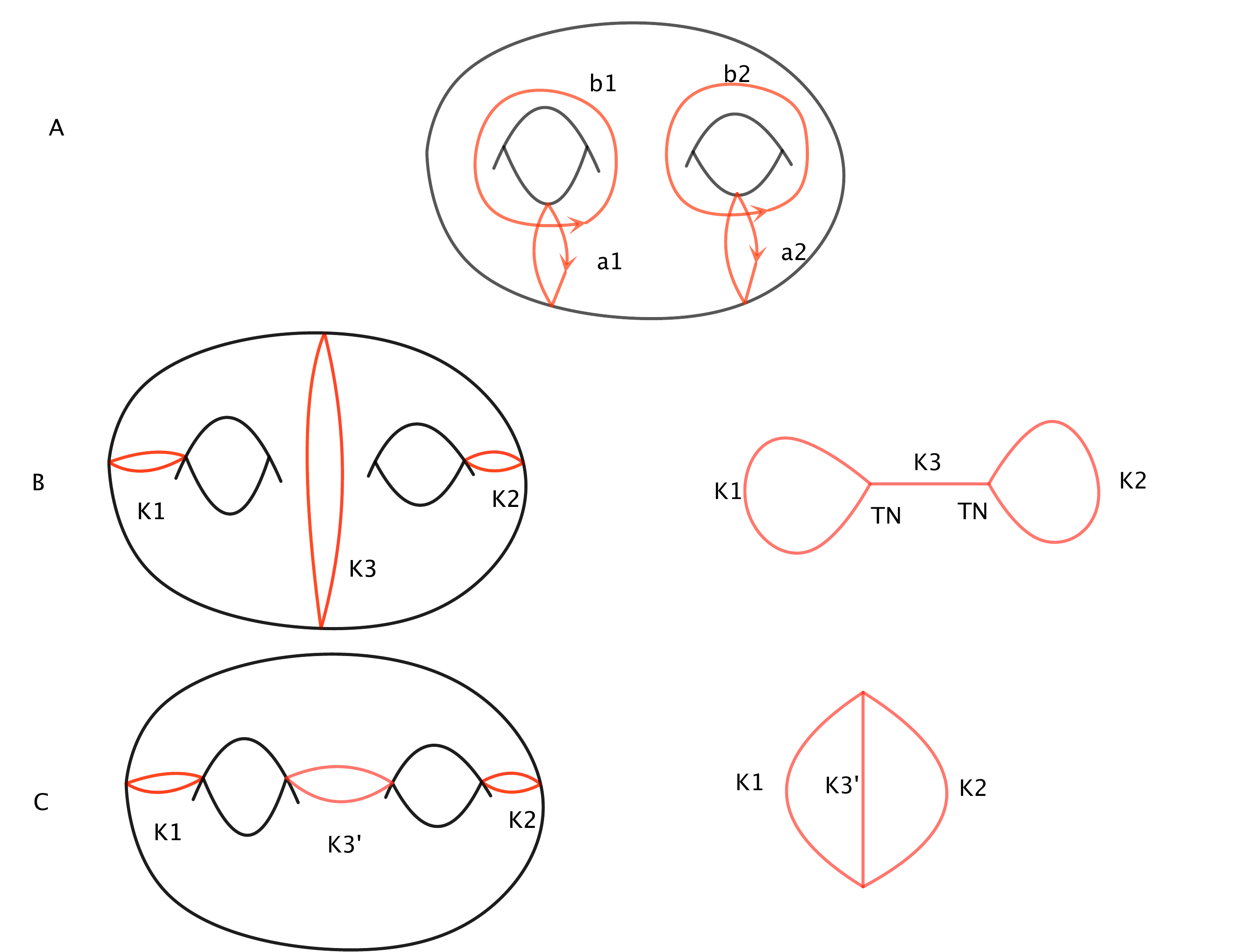}
\caption{Basis of homology class of genus two Riemann surface}
\label{TN}
\end{figure}
\end{center}

Now let's interpret the classification in terms of gauge theory. 
Let's start with a  weakly coupled duality frame represented by pants decomposition determined by three closed curves: $K_1, K_2, K_3$. This pants decomposition can be thought of first cutting the Riemann surface by $K_1$ and $K_2$, and get a fourth
punctured sphere which is further cut into two pants by $K_3$.  Here $K_1$ is in homology class $[a_1]$, and $K_2$ is in homology class $[a_2]$ and 
$K_3$ is in trivial homology class.  

According to our rules discussed earlier, the global form of gauge groups in the above duality frame are simple, and they are 
\begin{equation}
K_1:~~{SU(N)\over Z_{k_1}},~~~~~~K_2:~~{SU(N)\over Z_{k_2}},~~~~K_3:~~SU(N)
\end{equation}
We can do $T$ duality on $K_1$ and $K_2$ to find different sets of allowed set of line operators, and the details
is the same as the genus one case.  The $T$ transformation around $K_1$ change the homology class of line operators as 
\begin{align}
& T_{1}:~~n_1 k_1[a_1]+n_2 k_2[a_2]+p_1 k_1^{'}[b_1]+p_2 k_2^{'}[b_2]\rightarrow   (n_1 k_1+p_1 k_1^{'})[a_1] +n_2 k_2[a_2]+p_1 k_1^{'}[b_1]+p_2 k_2^{'}[b_2]       \nonumber \\
& T_{2}:~~n_1 k_1[a_1]+n_2 k_2[a_2]+p_1 k_1^{'}[b_1]+p_2 k_2^{'}[b_2] \rightarrow n_1 k_1[a_1]+(n_2 k_2+p_2 k_2^{'})[a_2]+p_1 k_1^{'}[b_1]+p_2 k_2^{'}[b_2]    \nonumber\\
& T_{3}:~~n_1 k_1[a_1]+n_2 k_2[a_2]+p_1 k_1^{'}[b_1]+p_2 k_2^{'}[b_2] \rightarrow n_1 k_1[a_1]+n_2 k_2[a_2]+p_1 k_1^{'}[b_1]+p_2 k_2^{'}[b_2] 
\end{align}

\begin{center}
\begin{figure}[htbp]
\small
\centering
\includegraphics[width=14cm]{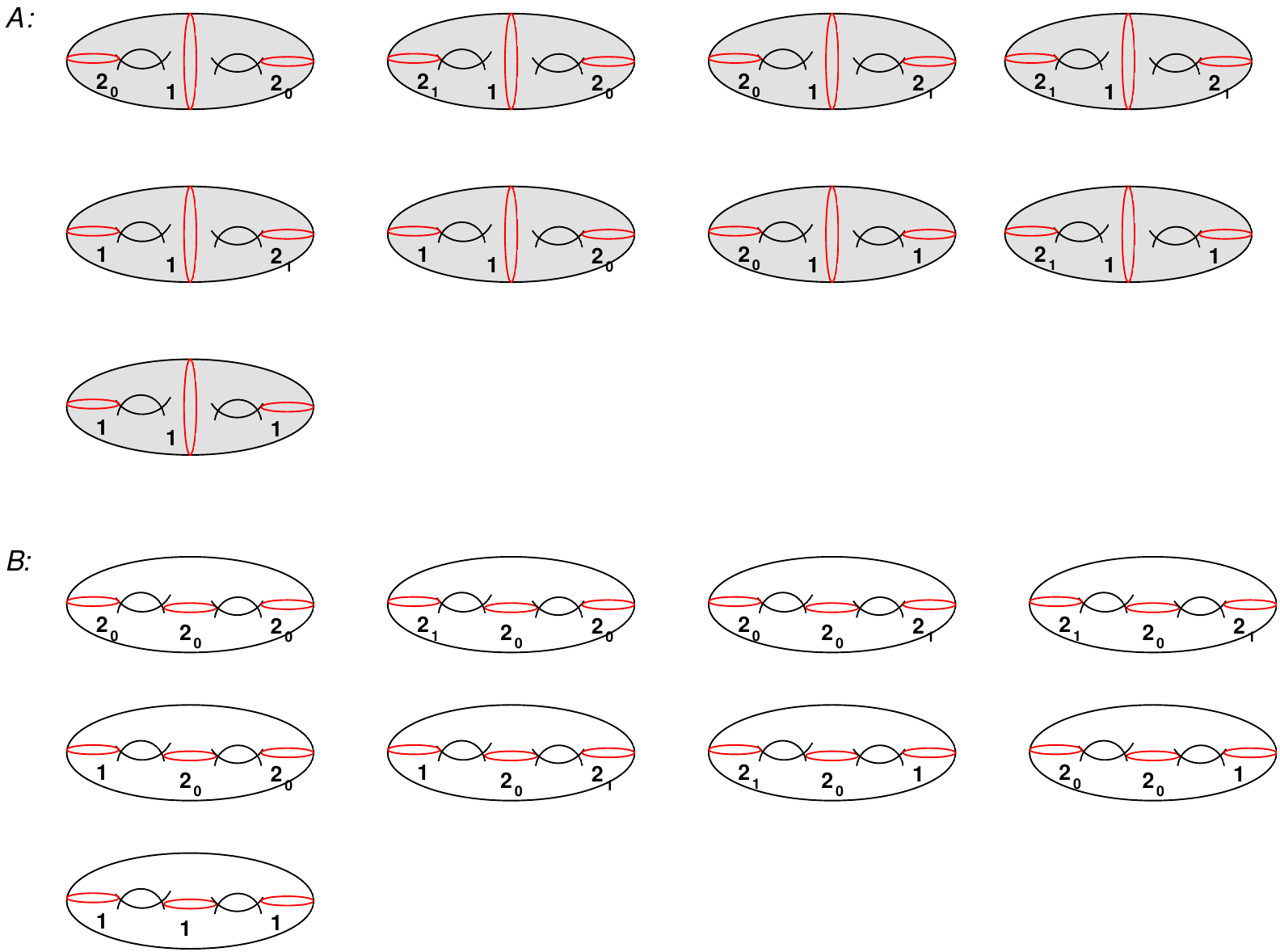}
\caption{Up: Different gauge theories associated with one duality frame. Bottom: Different gauge theories associated with another duality frame.}
\label{a1}
\end{figure}
\end{center}

\begin{center}
\begin{figure}[htbp]
\small
\centering
\includegraphics[width=12cm]{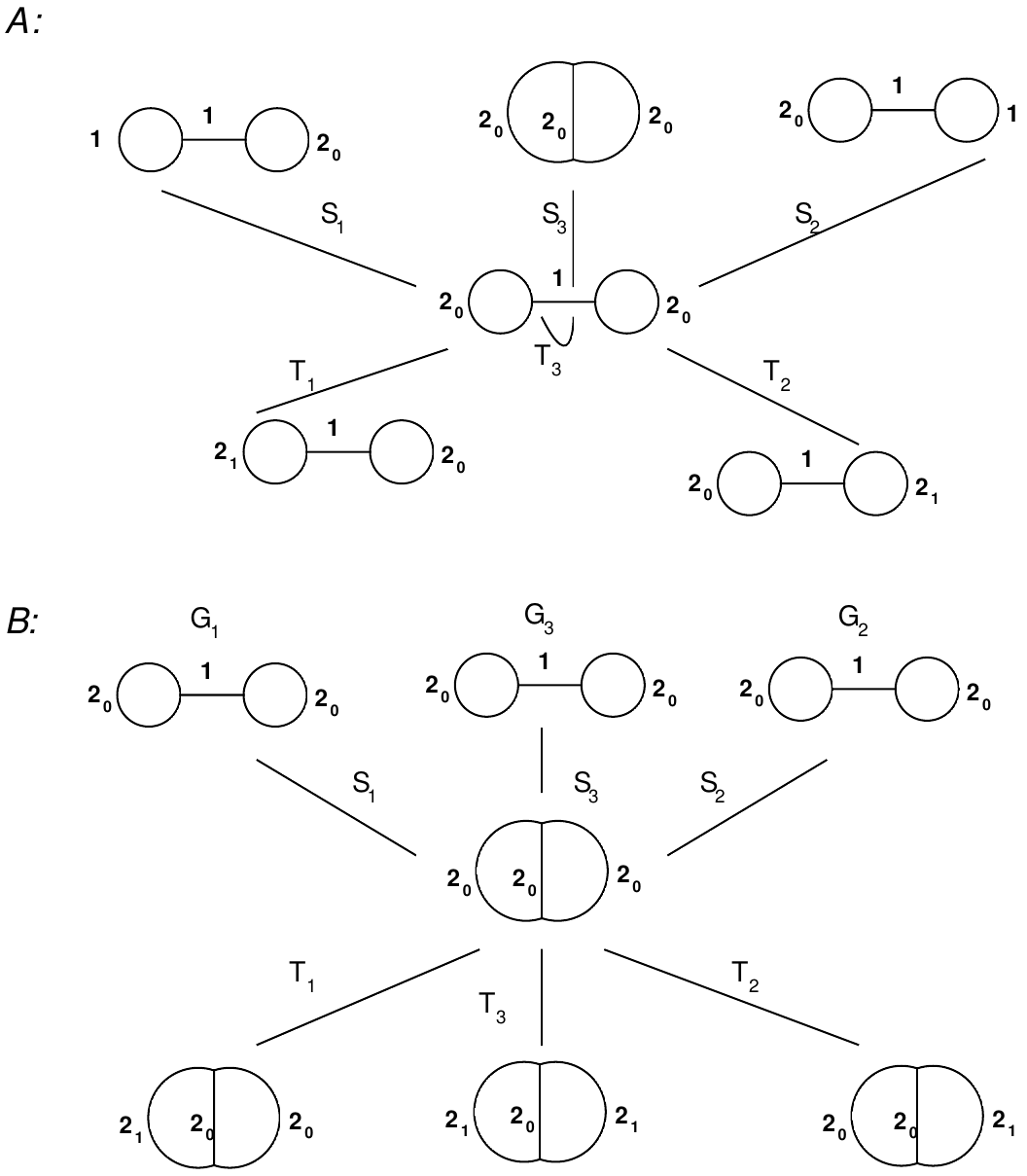}
\caption{Duality webs relating different gauge theories.}
\label{dual}
\end{figure}
\end{center}

What is interesting is to do S duality on $K_3$ and the dual theory is described by three closed curves $K_1,K_2,K_3^{'}$. 
Unlike $K_3$, $K_3^{'}$ is in non-trivial homology class $[a_1]-[a_2]$. Since we have chosen the minimal winding around $[a_1]$
and $[a_2]$, and we can find the gauge group around $K_3^{'}$ using the rules in last subsection. 
The $T$ transformation around $T_3$ is nontrivial now: 
\begin{align}
& T_{3}:~~n_1 k_1[a_1]+n_2 k_2[a_2]+p_1 k_1^{'}[b_1]+p_2 k_2^{'}[b_2] \rightarrow  \nonumber\\
&(n_1 k_1+p_1k_1^{'})[a_1]+(n_2 k_2-p_2k_2^{'})[a_2]+p_1 k_1^{'}[b_1]+p_2 k_2^{'}[b_2] 
\end{align}
However, it is easy to see $T_{3}=T_1 T_2^{-}$ and the Dehn twist around the circle $K_3^{'}$ will not generate new theories.  
So it is enough to consider $T$ transformation around $a_1$ and $a_2$. All possible choices of theories for $A_1$ theory is shown in
figure. \ref{a1}, and the duality webs relating these theories can be easily found using our formula for $S$ and $T$ transformation, see figure. \ref{dual}.

\newpage
\subsubsection{Arbitrary genus}
\begin{center}
\begin{figure}[htbp]
\small
\centering
\includegraphics[width=10cm]{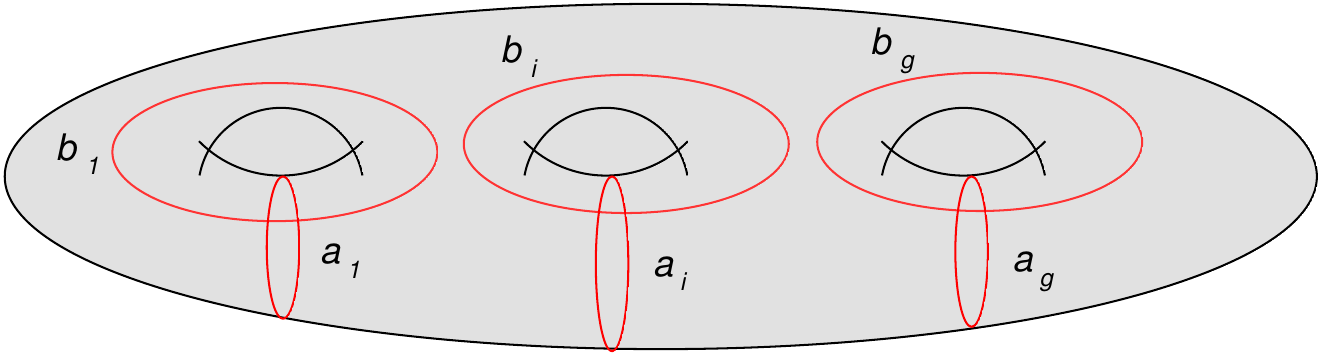}
\caption{Basis for homology group of a genus $g$ Riemann surface.}
\label{high}
\end{figure}
\end{center}

 For a genus $g$ Riemann surface without punctures (the puncture case does not introduce new features),
 the pants decomposition can be thought of as two steps: (1): Choose $g$ non-separating \footnote{By non-separating we mean the Riemann surface after cutting around $c_i$ is still connected.} closed curves $c_i$ to cut
 the Riemann surface into a punctured sphere; (2): Choose a pants decomposition of punctured sphere. It is easy to see that $c_i$ is a maximal set of commuting 
 basis for the homology class.  See figure. \ref{high} for a basis of homology group, and one can choose $a_i$ as a maximal set of commuting non-separating cut system. 

 To classify the line operators, we do the following: (1): Choose the minimal line operators around $c_i$; (2) The homology class of pure magnetic line 
 operators (by pure magnetic line operators we mean line operators in homology class generated by basis not included in $(1)$)  by imposing locality condition with the homology class chosen in (1); (3) Find the mixed 
 line operator imposing the locality condition with sets (1) and (2). Again, there are only $g$ discrete ndependent $\theta$ angles. 
 
The above choices can be regarded as taking zero $\theta$ angle for all the gauge groups. To find the allowed line operators for non-zero $\theta$ angles, we simply 
do $T$ transformation on various gauge groups, which will typically mix the electric and magnetic homology classes. The choice of line operators will be very rich. 

One can use our general formula of $T$ and $S$ transformation to relate different kinds of theories, and we leave the details to interested reader.

\section{Conclusion}
We studied various aspects of line operators of class ${\cal S}$ theory: we identify the geometric representations of Wilson-'t Hooft line operators and 
study the duality actions on them. We define the mutual locality condition on line operators. Then we use closure of OPE, mutual locality and 
maximality conditions to classify the allowed set of line operators. Finally we study the duality actions relating different gauge theories. The 
geometrical construction of line operators plays a crucial role in our applications.

There is one interesting lesson we learn about the choice of discrete $\theta$ angles:   one can not choose them independently and  
there are intricate relations among the choice of $\theta$ angles as we show for genus two example.
This interesting feature is related to the property of duality group or mapping class group. 
The topology of conformal manifold is entirely encoded in 
the property of mapping class group, and it does have important  effects on the choice of discrete $\theta$ angles. It is interesting 
to further study the mapping class group (duality group) using line operators.

We only consider 4d theory derived using 6d $A_{N-1}$ theory with full punctures, and it is interesting to generalize the study to more general case, i.e theory defined 
using non-full punctures. It is also interesting to generalize the consideration to D and E type theories.

We look at the classification of line operators using the geometric objects on the Riemann surface. The main reason is that these geometric 
objects are natural space on which duality action acts. We do not touch too much on four dimensional gauge theory meaning except the 
interpretation of discrete $\theta$ angle. It is pointed out in \cite{Aharony:2013hda}  that such choice of line operators has important implication for the physical theory derived 
by compactifying four dimensional theory on a circle (see \cite{Razamat:2013opa} for further study using index calculation.). What we want to point out is that the line operators in our picture also related to the 
four dimensional theory on a circle. In fact, the line operator in our case describes the canonical basis on the moduli space of Hitchin equation which 
actually describes the Coulomb branch of four dimensional theory reduced on a circle. Given different choices of line operators, we actually get 
different moduli space, therefore our choice of line operators indeed reflects the vacuum structure of four dimensional theory on a circle.

Moreover, these line operators are related to the Hamiltonian of the underlying Hitchin integrable system, and the quantization of this integrable 
system is related to the Nekrasov partition function of the gauge theory. It is interesting to see how the choice of line operators would affect the 
partition functions.
 
Similar 6d construction for a large class of $\mathcal{N}=1$ theories is presented in \cite{Xie:2013gma}, and we  have a generalized 
Hitchin equation in that context. It is natural to think of line operators of $\mathcal{N}=1$ theory in terms of closed curves on Riemann surface, and define its expectation value in terms of 
monodromy of commuting flat connections derived from generalized Hitchin equation.  
Although there is no BPS line operators of $\mathcal{N}=1$ theory, it is still possible to use the geometric construction to learn interesting dynamics of 
$\mathcal{N}=1$ theory, and we would like to report the progress in that direction in the near future.

\begin{flushleft}
\textbf{Acknowledgments}
\end{flushleft}
We thank Vasily Pestun, Shlomo Razamat, Yuji Tachikawa, Nathan Seiberg, Brian Willet,   Kazuya Yonekura and Peng Zhao  for helpful discussions.
This research is supported in part by Zurich Financial services membership and by the U.S. Department of Energy, grant DE-SC0009988  (DX).

\appendix{}
\section{A proof that mutual locality condition is independent of coordinate system}
The Dirac product between two line operators $L_1$ and $L_2$ are  defined as the coefficient of the leading order term in Poisson bracket of two line operators. 
Let's denote the tropical $a$ coordinates of $L_1$ as $a_{1i}$ and tropical $x$ coordinates of $L_2$ as $x_{2i}$. An important fact is that the dual $x$ coordinates are always
integer.  The Dirac product of two line operators is simply
\begin{equation}
<L_1, L_2>=-\sum _k x_{1k} a_{2k}.
\end{equation}
Let's do a mutation on a a quiver node $i$, then after the mutation, the tropical $a$ coordinates become
\begin{equation}
a_i^{'}=-a_i+\text{max}([{{\epsilon_{ij}}]_{+}a_j, [{\epsilon_{ij}}]_{-}a_j}),
\end{equation}
here $\epsilon_{ij}$ is the antisymmetric tensor of the quiver;
and the tropical $x$ coordinates become
\begin{align}
& x_i^{'}=-x_i \nonumber\\
& x_j^{'}=x_j+\epsilon_{ji} \text{max}(0, \text{sgn}(\epsilon_{ji})x_i)
\end{align}
The new intersection number is then
\begin{equation}
<L_1,L_2>=-\sum _k x_{1k}^{'} a_{2k}^{'}=-\sum x_{1k} a_{2k} +(\pm x_{1i}x_{2i}~\text{or}~0).
\end{equation}
since the dual $x$ coordinates are always integer, the Dirac product has the same fraction number behavior in new coordinate system. Therefore
if two line operators are mutually local in one coordinate system, they will be mutually local in any coordinate system related by cluster transformation. 

\bibliographystyle{utphys} 
 \bibliography{PLforRS}    
\end{document}